\documentclass[pdflatex,sn-mathphys-num]{sn-jnl}

\usepackage{graphicx}%
\usepackage{multirow}%
\usepackage{amsmath,amssymb,amsfonts}%
\usepackage{amsthm}%
\usepackage{mathrsfs}%
\usepackage[title]{appendix}%
\usepackage{xcolor}%
\usepackage{textcomp}%
\usepackage{manyfoot}%
\usepackage{booktabs}%
\usepackage{algorithm}%
\usepackage{algorithmicx}%
\usepackage{algpseudocode}%
\usepackage{listings}%
\usepackage{float}%
\usepackage{graphicx}   
\usepackage{subcaption} 
\usepackage{bm} 

\theoremstyle{thmstyleone}%
\theoremstyle{thmstyletwo}%

\theoremstyle{thmstylethree}%

\begin{document}

\title[Article Title]{A computationally efficient approach for predicting the transport properties of transition-metal alloys at elevated temperatures}


\author*[1]{\fnm{Akshay} \sur{Korpe}}
\author[4]{\fnm{Manish} \sur{Sudan}}
\author[5]{Ishtiaque K. Robin}
\author[4]{\fnm{Thomas} \sur{Berfield}}
\author[1 3]{\fnm{Garrett} \sur{Pataky}}
\author[4]{\fnm{Bikram} \sur{Bhatia}}
\author[5]{Osman El-Atwani}
\author[1 3]{\fnm{Enrique Martinez}}

\affil*[1]{\orgdiv{School of Mechanical and Automotive Engineering}, \orgname{Clemson University}, \orgaddress{ \city{Clemson}, \postcode{29634}, \state{South Carolina}, \country{United States of America}}}

\affil[2]{\orgdiv{Department of Material Science and Engineering}, \orgname{Clemson University}, \orgaddress{ \city{Clemson}, \postcode{29634}, \state{South Carolina}, \country{United States of America}}}

\affil[3]{\orgdiv{Department of Mechanical and Automotive Engineering}, \orgname{Clemson University}, \orgaddress{ \city{Clemson}, \postcode{29634}, \state{South Carolina}, \country{United States of America}}}

\affil[4]{\orgdiv{Department of Mechanical Engineering}, \orgname{University of Louisville}, \orgaddress{ \city{Louisville}, \postcode{40292}, \state{Kentucky}, \country{United States of America}}}

\affil[5]{\orgname{Pacific Northwest National Laboratory}, \orgaddress{ \city{Richland}, \postcode{29634}, \state{Washington State}, \country{United States of America}}}


\abstract{A novel phenomenological framework for an efficient estimation of the thermo-electric properties at room temperature and elevated temperatures of body-centered cubic (BCC) transition metal concentrated alloys is proposed in this work. The methodology is used to predict the electrical resistivity of BCC systems with our predictions showing excellent correlation with experimental data. This framework is further extended to predict the electrical resistivity $\mathrm{\rho}$, thermal conductivity $\mathrm{\kappa}$ and the specific heat capacity $\mathrm{C_{p}}$ of BCC alloys in the temperature range of 300 - 1300 K and the results are validated against experimental data. We demonstrate the capabilities of this model by using it to predict the thermo-electric properties of a concentrated $\mathrm{W_{53}Ta_{42}V_{5}}$ alloy which shows a saturation in the electrical resistivity $\mathrm{\rho}$ in the temperature range 300K-1300K. This model is then used to predict the properties of another concentrated $\mathrm{Nb_{40}Mo_{40}Ta_{20}}$  alloy in the same temperature regime.}

\keywords{Transport properties, conductivity, resistivity, Density-functional theory, bcc refractory metals, transition metals}

\maketitle

\section{Introduction}\label{sec1}
Transition metals are a group of elements that occupy the middle of the periodic table, ranging from groups III through XII. They are characterized by their partially filled d orbitals, and they are known to exhibit a wide range of mechanical and thermal properties. A special subset within this family are the refractory alloys which are known for their peculiarly high melting temperatures and high temperature mechanical performance, which make them ideal candidates for applications involving elevated temperatures. Furthermore, the advent of modern additive manufacturing methods, like laser powder bed fusion (LPBF) or electron beam melting (EBM), have made the practical viability of refractory alloys higher than ever before. Combining the vast compositional space of concentrated refractory alloys with the fact that their material properties are almost always drastically different than the weighted average of the constituent elements paves way for endless possibilities in the current landscape of material science. However, this emerging frontier presents its own set of challenges. The high melting temperatures and cost of raw materials make traditional synthesis and characterization approaches increasingly non-viable, highlighting the need for reliable computational frameworks to complement or even replace them. \par
This is especially true for refractory alloy design for thermal applications, which would require a good understanding of the ductility and thermo-electric properties at room temperature (RT) as well as at elevated temperatures. Although several computational models based on first principles exist, most are computationally expensive and are not suited for high-throughput mapping of the vast composition space of more complex alloys. Recently, ductility prediction has garnered more interest and a wide range of efficient semi-empirical models have been developed for these systems \cite{HU2021116800,KORPE2025102519,MAK2021104389,SINGH2023119104,LI2020174}. However, predicting the thermo-electric properties of these alloys efficiently remains elusive. Most popular approaches include the use of software packages like BoltzTrap2\cite{BoltzTraP2} for solving the Boltzmann transport equation (BTE) with a fixed relaxation time approximation. These results are then coupled with the actual values of relaxation time as a function of temperature, again using commercial packages such as EPW\cite{NOFFSINGER20102140} to compute the electron-phonon coupling coefficients. This method is quite robust for predicting the transport properties of pure elements and binary alloys but quickly becomes computationally prohibitive for complex systems due to the large super-cells as well as fine $k$ and $q$ meshes required for accurate results. \par

In this work, we propose a semi-empirical approach for predicting the thermo-electric properties (electrical resistivity, thermal conductivity and specific heat) of transition-metal concentrated alloys at temperatures ranging from RT to high temperatures. It is a two-step framework which involves computing the electrical resistivity of transition-element alloys using an empirical method and then coupling that with the solutions to the BTE with a fixed relaxation time approximation to obtain the electronic relaxation time as a function of temperature. This is then coupled with lattice properties such as the specific heat capacity $\mathrm{C_{p}}$ and the lattice thermal conductivity by solving the phonon BTE numerically. \par

The empirical framework for computing the electrical resistivity as a function of temperature is based on a semi-analytical relation between the probability of electron scattering into the unoccupied d-orbitals (unique to transition element alloys), which controls the transport properties in transition metals. These predicted values are coupled with the BoltzTrap2 package to extract the relaxation time and compute the electronic part of the thermal conductivity $\mathrm{\kappa_{e}}$. For the lattice contribution to thermal conductivity $\mathrm{\kappa_{l}}$ and the specific heat capacity $\mathrm{C_{p}}$, we use the VASP and Phonopy packages to compute the force constants and ShengBTE package to solve the phonon BTE using numerical methods. The predictions are validated using experimental data and the results are discussed.\par

\section{Theoretical framework}

\subsection{Resistivity of transition metals}
According to Mattheissen's rule, the resistivity of any pure metal can be attributed to two contributions in the form :
\begin{equation}
\label{eq1}
    \rho = \rho_{r} + \rho_{i}(T)
\end{equation}
Where, the $\mathrm{\rho_{r}}$ is the residual resistivity and is a constant that is not a function of temperature and $\mathrm{\rho_{i}}$ is the ideal resistivity which depends upon the scattering of conduction electrons. According to Debye theory, the phonons are no longer quantized for temperatures larger than the Debye temperature $\mathrm{\Theta}$ in metals and the rate of electron scattering due to phonons becomes proportional to the square of the amplitude of the ions about their mean position, which makes $\mathrm{\rho_{i}}$ proportional to temperature (T)\cite{ziman1967electrons}. Using this, $\mathrm{\rho_{i}}$ for metals at high temperatures follows the functional form:

\begin{equation}
\label{eq0}
    \rho_{i} = KT/\Theta^{2}
\end{equation}

Where K is a constant and $\mathrm{\Theta}$ is the characteristic Debye temperature for a material. This however, does not hold true for transition metals, which deviate from this linear behavior. Interestingly, there is a pattern observed in these deviations for refractory metals. The resistivity of group III and V elements increases slower than linear and for group IV and VI the opposite is observed. Chiu et al.\cite{PhysRevB.13.1507} have beautifully captured this effect experimentally by studying the normalized resistivity vs temperature for different alloying percentages for binary refractory alloys. This peculiar thermo-electric behavior in transition metals has been attributed to the presence of vacant d-orbitals. In this work, we use the relation between d-orbitals and resistivity and establish a framework to accurately account for this behavior.\par

Equation \eqref{eq1} has two independent components that constitute the total electrical resistivity at a given temperature. Since $\mathrm{\rho_{r}}$ is independent of temperature, the only way to obtain its values is through extrapolation of cryogenic experiments. For the sake of this work, we will assume that pure elements are ideal metals and neglect the term. This assumption is valid since for most metals, this value is very small as compared to the values at the temperature scales that we are concerned with \cite{Desai10.1063/1.555723}. For the temperature dependent term $\mathrm{\rho_{i}}(T)$, we use the theory for transition metals by Mott and Jones\cite{P1937,PhysRevB.13.1507}. The total resistivity $\mathrm{\rho}$ for $T>\mathrm{\Theta}$ is given by the equation:
\begin{equation}
    \label{eq2}
    \begin{split}
    \rho=\left (\frac{K}{\Theta_{0}^2} \right )T\left (1+6\alpha \gamma T \right ) \left (1-AT^{2} \right ) \\
    A =\frac{\left(\pi k_{b}\right )^{2}}{6}\left [3 \left (\frac{1}{\nu}\frac{d\nu}{ d\epsilon} \right )^{2}-\frac{1}{\nu}\frac{d^2 \nu}{d\epsilon^2} \right ] _{\epsilon_{F}} 
    \end{split} 
\end{equation}
Where, $K$ is a constant of proportionality, $\mathrm{\Theta_{0}}$ is the Debye temperature at 0 K, $\mathrm{\alpha}$ is the linear thermal expansion coefficient and $\mathrm{\gamma}$ is the Gruneisen constant. The cumulative term $\mathrm{\left ( K/{\Theta_{0}^2} \right ) \left (1+6\alpha \gamma T\right )}$ is the characteristic Debye temperature $\mathrm{\Theta_{D}}$ for the metal and encapsulates the change in $\mathrm{\Theta}$ with temperature. As discussed in later sections, we assume that this $\mathrm{\Theta_{D}}$ is a constant term in this work. This leaves us with an equation for $\mathrm{\rho}$ as a function of $\mathrm{\rho_{i}}$ times the $(1-A\mathrm{T^{2}})$ term. $\mathrm{\rho_{i}}$ is a linear function in T and A is related to the probability that a conduction electron will make a transition from an s orbital to a d orbital via scattering, which correlates to the electronic density of states $\mathrm{\nu}$ and the Fermi energy $\mathrm{\epsilon_{F}}$. The s and p orbitals are mainly responsible for the conduction and the localized d-orbitals above the $\mathrm{\epsilon_{F}}$ can be thought to act as traps for the conduction electrons and hence result in increased electrical resistivity. A rigorous study for Nb and Mo that validates this theory, albeit for low temperatures, was done by Yamashita et al.\cite{10.1143/PTP.51.317b}. While the methods used in that work may be less suitable for multi-component concentrated alloys, particularly at high temperatures due to computational complexity, the results nonetheless reinforce the underlying theory and motivate further research in this direction.\par

For the first part of this work, we use a functional form similar to that of $\mathrm{\rho}$ from Eq. \ref{eq2}  using the assumptions stated previously. The resistivity is hence computed using the equation:
\begin{equation}
    \label{eq3}
    \rho = \rho_{i}(1-AT^{2})
\end{equation}

We hypothesize that $A$ is a material property unique to an alloy composition, and we compute it for pure W, Ta, Nb and Mo by plugging in experimental electrical resistivity values for high temperatures in Eq. \eqref{eq3}. The values of $\mathrm{\rho}$ are normalized at 300 K for these elements and plotted against an ideal reference line as predicted by the Bloch-Gruneisen model. The deviations from the linear reference are then attributed to the $\left ( 1-A\mathrm{T^{2}} \right)$ term and values of $A$ are computed. Figure \ref{figA_pure_elements} shows the normalized resistivity vs temperature for W, Ta, Nb and Mo from 300 K to 1400 K. As expected from the previously mentioned works and theory, W and Mo (group VI) elements deviate positively and Nb and Ta (group V) deviate negatively from the ideal linear reference. We can interpret this as some metals increase resistivity with temperature faster than ideally while others do it much slower. We postulate that when these transition metals form alloys, the change in $\mathrm{\rho}$ is dominated by s-d scattering since that is the dominant mode of scattering in this class of metals\cite{ziman1967electrons}. Since we use normalized datasets, the values of $\mathrm{\rho_{i}}$ should be the same for all pure metals as well as alloys formed by transition elements and the effects of alloying on resistivity should be captured entirely by the change in $A$. \par

Since the model so far has been set up only in the form of normalized resistivity, it only helps us to predict the trends at elevated temperatures relative to a normalized initial value rather than the absolute values. For a fully predictive model, one would need to obtain at least one absolute value at any given temperature. Normally this is not an issue for pure metals or common alloys for which we have RT resistivity data available. But in process of optimizing the compositional space of concentrated alloys, obtaining data becomes practically infeasible due to the difficulty and cost of manufacturing samples for experimental measurements. This combined with the observation that the properties of alloys are highly sensitive to the elemental concentrations~\cite{KAO20111607,article} highlights the need for a method to predict the resistivity of complex refractory alloys. \par

\subsection{Resistivity of alloys}
It is a general observation that the resistivity of alloys is always greater than the expected weighted average of the constituent elements. We hypothesize that this increase can be quantified only by the change in the electronic band structure, crystal structure and the lattice motif. If the crystal structure of the alloy remains the same as that of the constituent metals, the only difference between an alloy and a pure system is the local deviations in the electronic environment due to the presence of different atomic species in the alloy. These fluctuations are hypothesized to cause an additional `friction' to the flow of electrons in the conduction process. This is related to the elastic scattering caused by alloying and is not temperature dependent. To account for this effect, we propose the definition of $\mathrm{\rho}$ for alloys as follows:
\begin{equation}
    \label{eq4}
    \rho = \rho_{o} + \rho_{r}^{alloy}
\end{equation}

\begin{figure}[htbp]
\centering
\begin{minipage}{0.48\linewidth}
    \includegraphics[width=\linewidth]{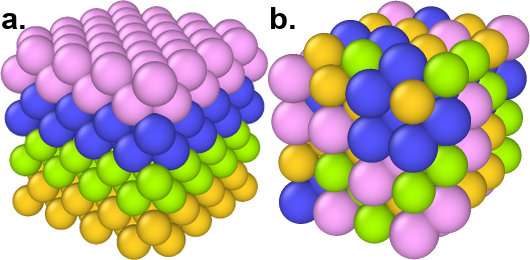}
\end{minipage}%
\hfill
\begin{minipage}{0.48\linewidth}
    \caption{\textbf{Mixing of metals to form an alloy with the same crystal structure.} }
    \label{figcombined}
    \textbf{a)} Layering stacking. \textbf{b)} Disordered mixing.
\end{minipage}
\end{figure}

The $\mathrm{\rho_{o}}$ is coined to be the alloy resistivity with `ideal mixing' and the constant term $\mathrm{\rho_{r}^{alloy}}$ is hypothesized to be related to the disorder of the system, meaning it is zero for pure metals without any impurities. In this work, for the sake of simplicity, we only consider single phase solid solutions and assume no short-range order (SRO). To quantify this disorder, we use the ideal configurational entropy per atom for solid solutions as a surrogate parameter given by \cite{yeh_article}:
\begin{equation}
    \label{eq5}
    S_{conf} = \sum_{i} \chi_{i} ln(\chi_{i})
\end{equation}
Here, $\mathrm{\chi}$ is the concentration of the elements in the alloy and $S_{conf}$ is the ideal configurational entropy. To better illustrate this proposed concept, Fig. \ref{figcombined}-a represents what we call `ideal mixing' ($\mathrm{\rho_{o}}$) where we see that layers of different atom types are stuck on top of each other and hence, in this case, provided the effect of the interface between layers is negligible, the resistivity of the alloy would just be the weighted average of the constituent elements and $\mathrm{\rho_{r}^{alloy}}$ is zero. We say that this case can just be treated as a pure bulk metal and equation \ref{eq1} is valid. Figure \ref{figcombined}-b shows the case which is closer to body-centered cubic (BCC) alloys, where there is a dispersion of the constituent elements and we see disorder in the system. According to our theory, this will have an increased resistivity as compared to the ideal case and the additional value is represented by the $\mathrm{\rho_{r}^{alloy}}$ term. We use available experimental data for pure elements, dilute alloys and concentrated alloys at RT and compute the values of $\mathrm{\rho_{r}^{alloy}}$ using the equation:
\begin{equation}
    \label{eq6}
    \rho_{r}^{alloy} = \rho(T) - \rho_{o}(T)
\end{equation}
Here, $\mathrm{\rho_{o}}$ is the weighted average of the resistivity of the constituent alloying elements and $\mathrm{\rho}$ is the total experimental resistivity. These values are then fitted against $\mathrm{S_{conf}}$ to obtain a predictive curve as discussed previously. We can then use Eq. \ref{eq2} to compute the absolute values of resistivity at higher temperatures. We do this for alloys with experimental data available across the temperature range of 300-1300 K and find excellent correlation, as shown in Fig.~\ref{figHEA_side_by_side}.\par

\begin{figure}[H]
    \centering
    \includegraphics[width=1.05\textwidth]{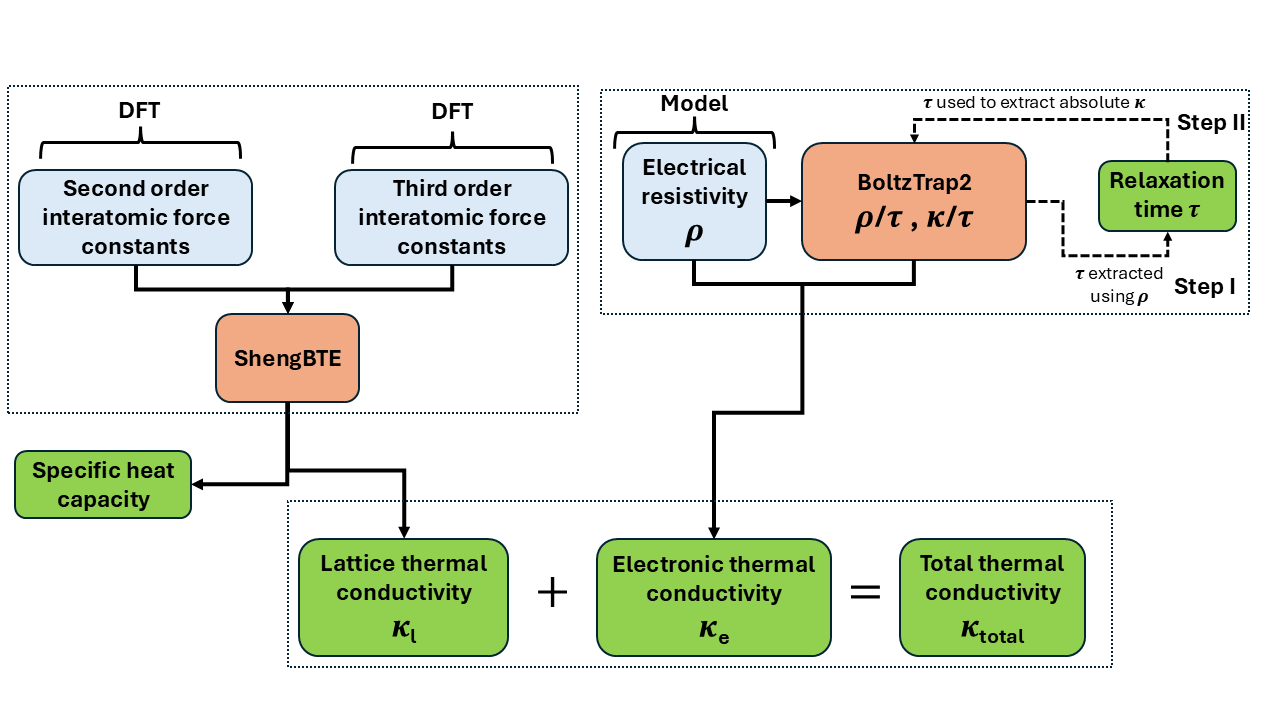} 
    \caption{Workflow for computing the total thermal conductivity. The blue blocks are the parameters that are computed using DFT and the theory developed in the manuscript, the orange blocks are numerical solvers for the Boltzmann transport equation and the green blocks represent predicted values.}
    \label{flow chart}
\end{figure}

Now that we have established a complete phenomenological framework for predicting the absolute values of resistivity for alloys, we make use of this to compute the thermal conductivity. We make use of the BoltzTrap2 package \cite{BoltzTraP2} to compute the thermo-electric properties using a constant relaxation time approximation and then use the predicted resistivity values from our model to extract the relaxation times as a function of temperature. Using this approach we calculate the values of the electronic component of the thermal conductivity $\mathrm{\kappa_{e}}$ from the BoltzTrap2 results. For the lattice properties, like the lattice part of thermal conductivity $\mathrm{\kappa_{l}}$ and the specific heat capacity $\mathrm{C_{p}}$, we use the ShengBTE \cite{ShengBTE_2014} package to numerically solve the BTE for phonons. We thus obtain the electrical resistivity $\mathrm{\rho}$, total thermal conductivity $\mathrm{\kappa(T)}$ = $\mathrm{\kappa_{e}}$ + $\mathrm{\kappa_{l}}$ and specific heat $\mathrm{C_{p}}$. The results of our model are compared with experimental data. Figure \ref{flow chart} shows the workflow for computing the total thermal conductivity.\par
Finally, using the framework developed, we study two concentrated ternary alloys, $\mathrm{W_{53}Ta_{42}V{05}}$ and $\mathrm{Nb_{40}Mo_{40}Ta{20}}$. The thermal conductivity and specific heat values of the W-based alloy are known experimentally and used for validation, while the properties of the Nb alloys are predicted in this work. The experimental validation is left for a future study, as the synthesis of the alloy is currently in progress.

\subsection{Conditions for the validity of this framework}
\label{checks}
\subsubsection{Resistivity saturation}
\label{resistivity sat}
Transition metals and their alloys are known to experience resistivity saturation at elevated temperatures when they reach their Ioffe-Regel limits \cite{loffe_regel_experimental_RevModPhys.75.1085,loffe-regel-thory-1-PhysRevB.24.7404}. The basic idea is that at some temperature, the mean free path of conduction electrons in these metals reaches values that are comparable to the interatomic distances, and around that temperature the electrical resistivity stagnates with no further increase with temperature. Experiments show that the resistivity for most pure transition metals follows an exponential growth curve and saturates at elevated temperatures but the saturation temperatures vastly vary. Generally, as the number of elements in the alloy increases, the saturation temperature significantly decreases  \cite{loffe-regel-alloy-1-1992674,loffe-regel-alloy-2-2SUNDQVIST2021127291}. It is hence very crucial to check for this saturation for the alloys being studied. If the alloy has reached resistivity saturation in the temperature regime under consideration, instead of using the resistivity from the previous framework, the saturated values need to be used.\par

For the sake of this study, since we already have our parameters set up in the form of the electronic relaxation time $\mathrm{\tau}$, we use that as a surrogate indicator of resistivity saturation, since that is linearly related to the electronic mean-free path. We define a parameter K given by:
\begin{equation}
    \label{eq:Kparam}
    \tau \propto T^{K}
\end{equation}
where $\mathrm{\tau}$ is the average electronic relaxation time and T is the temperature. It is proposed that if the value of K is closer to 0, the material has reached the Ioffe-Regel limit and the resistivity has stagnated. Here, we set a threshold of K=0.2 for a material to have reached the Ioffe-Regel limit. The threshold has not been set to 0 to accommodate computational error margins. \par

It has to be noted that since the mean-free paths and relaxation times are such elusive concepts, it is not possible to accurately determine their average values, especially if we want to keep the model efficient. Hence, in this work, we use experimental data to determine the values of relaxation times as discussed in previous sections. The model still remains useful to determine the effects of alloying on a previously studied alloy as we assume that minor tweaks to the alloying composition is not expected to change the Ioffe-Regel limit drastically. Also, the saturation behavior for alloys with data available at lower temperature can be estimated and the properties at elevated temperatures can then be predicted using this model.

\subsubsection{Violation of Wiedemann-Franz relation}

Although rare, the Wiedemann-Franz relation has been observed to be violated by materials. The Wiedemann-Franz law relates the electronic thermal conductivity $\mathrm{\kappa_{e}}$ and the electrical resistivity $\mathrm{\rho}$ with the temperature as follows:
\begin{equation}
    \kappa_{e} \rho = LT
\end{equation}
Where, L is the Lorenz number and is equal to $\mathrm{2.44x10^{-8} W\Omega K^{-2}}$ (Sommerfeld value). Although most common metals follow this relation, the value of L is not a universal constant and is known to fluctuate from the Sommerfeld value. Since this relation is so pivotal in the current framework, we need to verify its validity before we can use it. For this, we use the $\mathrm{\kappa_{e}/\tau}$ and $\mathrm{\sigma/\tau}$ ($\mathrm{\sigma}$ is the electrical conductivity equal to 1/$\mathrm{\rho}$) values from the Boltztrap2 output within a fixed relaxation time approximation (the $\mathrm{\tau}$ term cancels out in the ratio). Again, to accommodate computational error, we assume that the Wiedemann-Franz relation holds as long as the Lorenz number $L$ stays in the range of 2-2.6 in the working temperature regime as it is not expected to catastrophically affect the predicted thermo-electric values. \par

\subsection{Saturation of s-d scattering}
\label{s-d}
Since there is a finite DOS at any energy value, we would expect the effects of this scattering on the $\mathrm{\rho}$ to be finite and that it would stagnate to a constant value at elevated temperatures. This means that the $A$ tends to move towards a zero value at elevated temperatures as there are no new available DOS and an equilibrium state is achieved. Ideally, this transition is expected to be smooth and the $\mathrm{1-AT^{2}}$ term is expected to vanish at this point, or else the mathematical equation will yield nonphysical results like reducing or even negative $\mathrm{\rho}$ at elevated temperatures. As mentioned before, for this work, we will be treating $A$ as a constant, which means that we turn its value to zero at a cut-off temperature to prevent nonphysical results. to determine that cut-off T, we use the $\mathrm{\tau}$ vs T plot as we do for \ref{resistivity sat} and we ascribe the point of zero slope to be this temperature. So in this model, the value of A remains constant till this temperature and then turns to null above it. 

\section{Materials and methods}

 \subsection{DFT details}
The bulk super-cells for all alloys were created using the ATOMSK \cite{atomsk_HIREL2015212} package and the ground state relaxations were done using the VASP \cite{vasp_PhysRevB.54.11169} package at zero pressure. All DFT calculations in this work were conducted using 
GW-sv pseudopotentials based on projector-augmented wave method (POTPAW-GW-sv) with an energy cut-off for plane-wave basis of 500 eV. For the ground state ionic relaxation, supercells consisting of 72 atoms were used with a k-mesh size of $3\times 3 \times 3$. The energy tolerance in the minimization process was set to $\mathrm{10^{-4}}$ eV with gaussian smearing with a width of 0.2 eV. For the Self-Consistent Field (SCF) calculations, a finer k-mesh of $9\times 9 \times 9$ was used with a Methfessel-Paxton smearing and a smearing value of 0.1 eV. For the Non Self-Consistent Field (NSCF) calculations a mesh size of $15\times 15\times 15$ was used using tetrahedron Blöchl smearing with a smearing value of 0.05 eV. The analysis of the DOS was performed using the VASPKIT \cite{VASPKIT} package. The second-order force constants were computed for the same supercells using the phonopy \cite{phonopy-phono3py-JPCM} package and the thirdorder.py \cite{ShengBTE_2014} code was used for computing the third-order force constants. A $10\times 10\times 10$ q-mesh was used for the numerical solution of the BTE using the ShengBTE package.

\subsection{Stochastic parameter $A$}
\begin{figure}[htbp]
    \centering
    \includegraphics[width=0.6\textwidth]
    {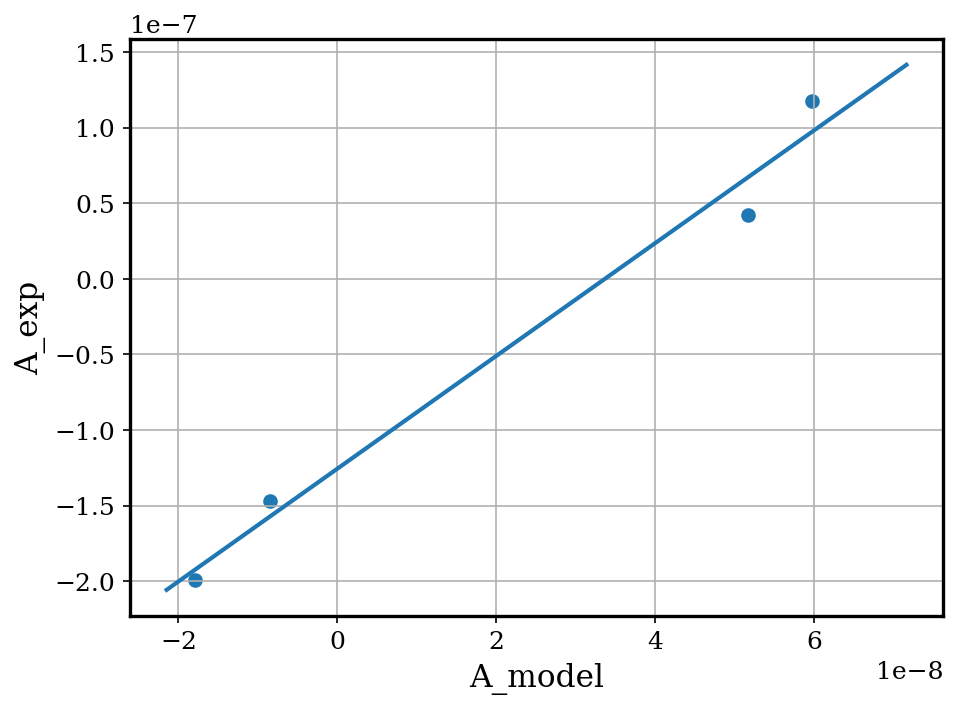}
    \caption{Correlation between scattering parameter A from experiments and A from our model.}
    \label{figA_DOS_pure}
\end{figure}

\begin{figure}[H]
    \centering

    \begin{subfigure}{0.46\textwidth}
        \centering
        \includegraphics[width=\linewidth]{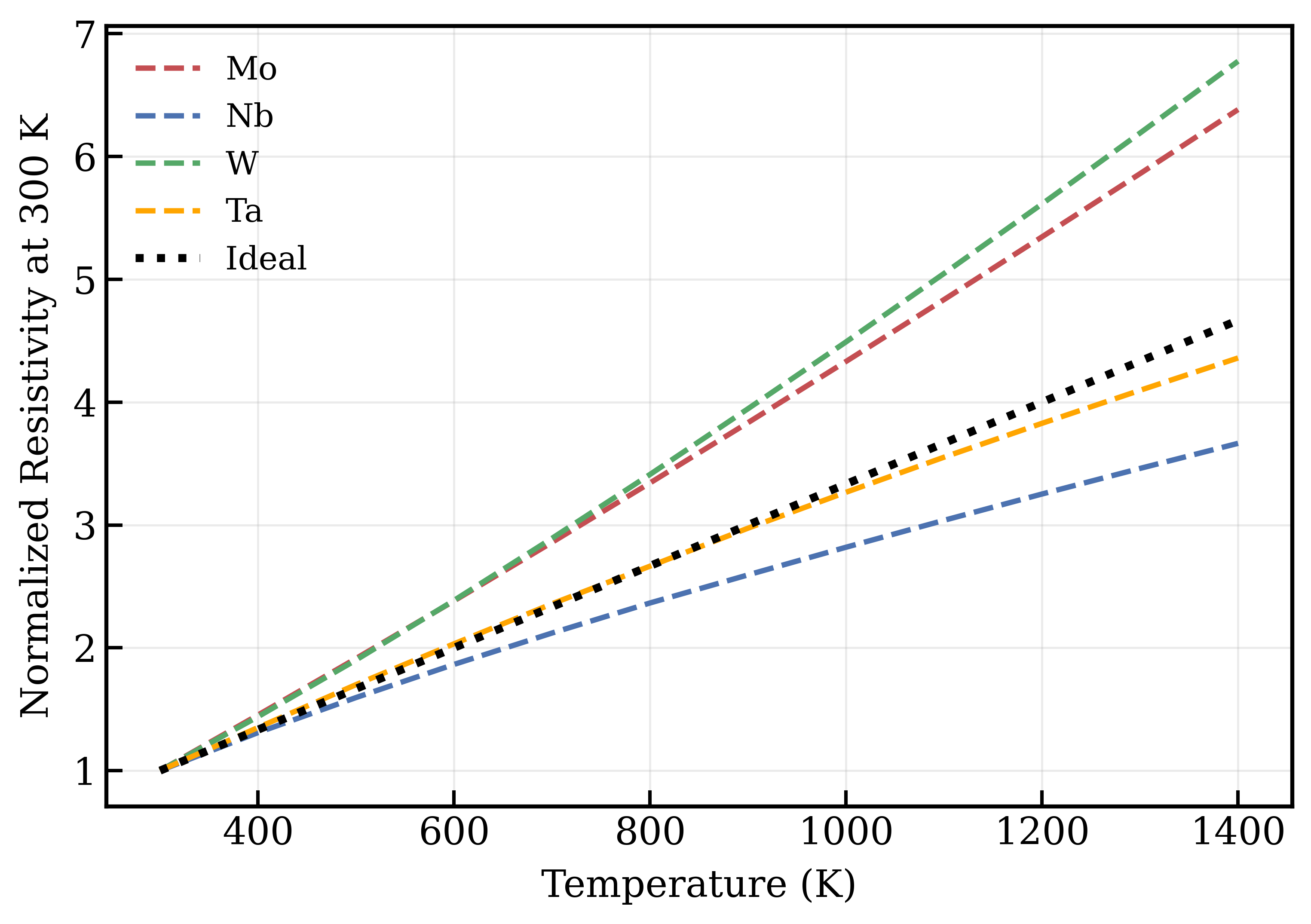}
        \caption{Ideal vs. experimental resistivity (normalized at 300 K).}
        \label{figpure_elements}
    \end{subfigure}
    \hfill
    \begin{subfigure}{0.46\textwidth}
        \centering
        \includegraphics[width=\linewidth]{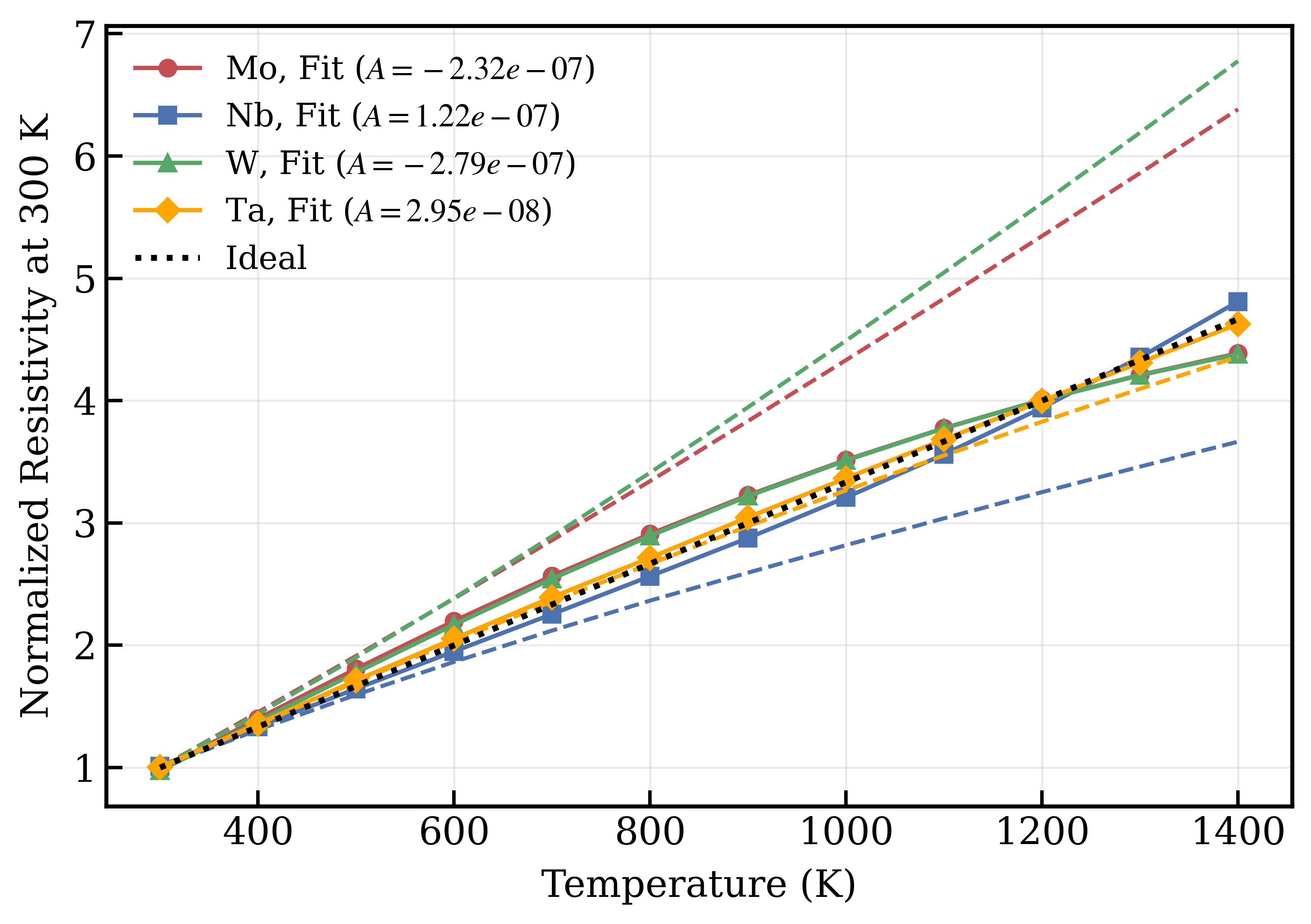}
        \caption{Extracted \( A \) values from normalized resistivity vs. temperature curve.}
        \label{figA_pure_elements}
    \end{subfigure}
    \caption{Comparison of resistivity behavior and scattering parameter A for pure BCC transition metals.}

    \label{figtwo_plots}
\end{figure}

The values of the stochastic parameter $A$ for W, Ta, Nb, Mo, $\mathrm{Nb_{90}Mo_{10}}$ and $\mathrm{Nb_{95}Mo_{05}}$ were computed from equation \ref{eq3} by using experimental data for the resistivity vs temperature relation available in previous works \cite{Desai10.1063/1.555723,Moore1980}. The experimental value of  $\mathrm{\rho_{i}}$ (which equals the difference between $\mathrm{\rho}$ from experiments and $\mathrm{\rho_{r}^{alloy}}$ from the model, as discussed next) are then normalized at 300 K (i.e $\mathrm{\rho / \rho(300K)}$) as shown in Fig. \ref{figpure_elements} and the values of $A$ are computed using the ideal linear reference line in the normalized plot. We can conveniently use room temperature data since transition metals in this study have $\mathrm{\Theta_{0}}$ values around this temperature. The results are shown in Fig. \ref{figA_pure_elements}. The obtained average $A$ values per atom are then plotted against the computed $A$ values using equation \ref{eq2} for each element at 0 K and a linear behavior is observed. The best linear fit is given by:
\begin{equation}
    \label{A_fit}
    A = 0.5A_{DFT} - 1.21 \times 10^{-7}
\end{equation}
Where, $\mathrm{A_{DFT}}$ is computed using DFT from equation \ref{eq2}. The reason for the $\mathrm{A_{DFT}}$ values not matching the experimental $A$ values is likely due to the fact that A is not a constant but changes slightly with temperature (as evident from figure \ref{figA_pure_elements} not being exactly linear) due to the changing coefficient of thermal expansion as well as the $\mathrm{\Theta_{D}}$ of these metals with temperature. However, the excellent statistical correlation by assuming constant $\mathrm{\Theta_{D}}$ and neglecting the effects of thermal expansion as seen in table \ref{tab:electrical_resistivity_prediction} shows that treating $A$ as constant in this temperature regime is a good approximation for predicting $\mathrm{\rho}$. Equation \ref{A_fit} predicts the $A$ parameter that controls the trends in electrical resistivity at elevated temperatures. Note that the plot in Fig. \ref{figA_DOS_pure} reverses sign at $\mathrm{A_{DFT}=2.42\times 10^{-7}}$ and by the nature of equation \ref{eq3}, the resistivity will increase faster for elements to the left of this value (negative) and vice-versa for elements to the right (positive), as compared to ideal. This is in accordance with the experimental observations that group VI elements (W and Mo) and group V elements (Nb and Ta) increase resistivity faster and slower than ideal (see Fig. \ref{figA_pure_elements}), respectively. \par

\subsubsection{Geometric approximation of the DOS plot}
The DFT methods have an inherent limitation due to the finite K-mesh sizes possible in order to optimize the efficiency and accuracy of the computation. This results in a noisy DOS plot which makes it difficult to interpret the raw DFT data. Our model uses an analytical expression containing the slopes and curvature of the DOS at $\mathrm{\epsilon_{F}}$, which makes it even more sensitive to local fluctuations and hence it is important to make approximations to smoothen the curve in order to be able to produce consistent and reproducible results. In this work, we have proposed a simple method to do so. Phenomenologically, we see that the $\mathrm{\epsilon_{F}}$ always lies in a region on the DOS between 2 high peaks and a trough in between on the global scale. We assume that these peaks and the trough have zero slopes and fit 2 curves containing one peak of either side and the trough. Since we have 4 boundary conditions on each side (2 zero slopes on the peak and trough and the points themselves), we are able to fit quartic curves on the left and right hand side of the trough that we use to compute the A parameter. The equations used to compute the A as well as the curves themselves can be seen in figure \ref{A_fit}. This method has been found to be in excellent correlation with the experimental A values and has been tested to be highly reproducible. This also is in line with the idea that alloys with $\mathrm{\epsilon_{F}}$ near the peaks (Nb and Ta) have the maximum DOS for D orbitals and hence they have the maximum probability of occupation which makes their resistivity just above the Debye temperature much higher than those that have their $\mathrm{\epsilon_{F}}$ near the trough which show the opposite (W and Mo).

\subsection{Alloy residual resistivity}
\begin{figure}[htbp]
    \centering
    \begin{subfigure}{0.48\linewidth}
        \centering
        \caption{}
        \includegraphics[width=\linewidth]{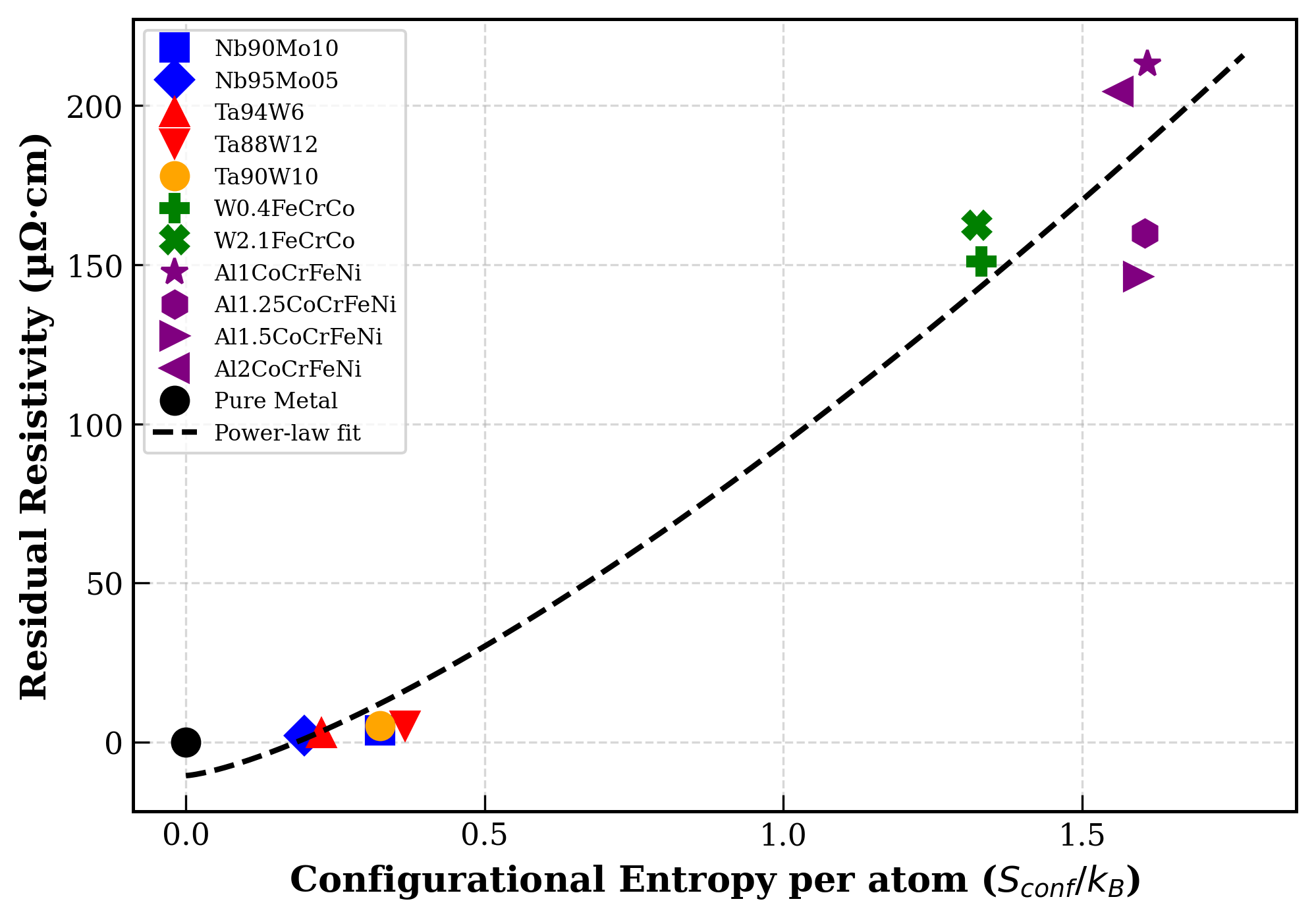}
        \label{figHEA_conf_entropy}
    \end{subfigure}\hfill
    \begin{subfigure}{0.48\linewidth}
        \centering
        \caption{}
        \includegraphics[width=\linewidth]{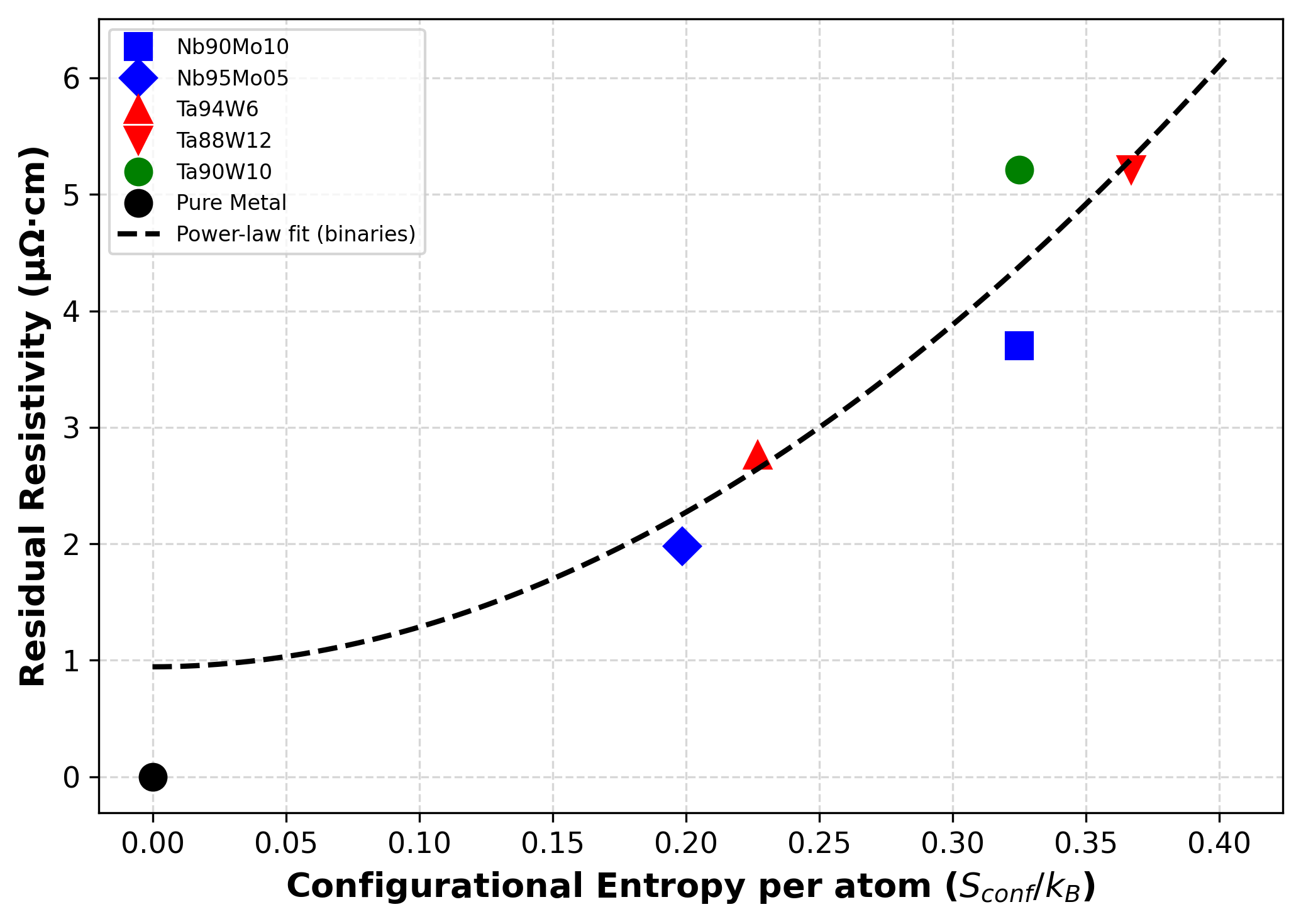}
        \label{figHEA_rho_residual}
    \end{subfigure}
    \caption{Configurational entropy per atom vs residual resistivity for single-phase BCC solid solutions.}
    \label{figHEA_side_by_side}
\end{figure}

The values of $\mathrm{\rho_{r}^{alloy}}$ at temperatures just above the Debye temperature can be computed using equation \ref{eq6}. Figure \ref{figHEA_conf_entropy} shows experimental $\mathrm{\rho_{r}^{alloy}}$ versus $\mathrm{S_{conf}}$ per atom for some single phase BCC solid solutions, where $\mathrm{\rho_{r}^{alloy}}$ is computed and substituting the $\rho$ term with experimental resistivity \cite{TAYLOR1971369,Moore1980_ThermalTransport_Nb,KAO20111607,article}. Using this fit, the equation of the curve that can be used to predict the $\mathrm{\rho_{r}^{alloy}}$ for any random BCC solid solution is given by:
\begin{equation}
\label{eq8}
    \rho_{r}^{alloy} = 104.20(S_{conf(per atom)}^{1.36}) - 10.56
\end{equation}
All data points are color coded to indicate different experimental works. We get an excellent fit with a Pearson correlation coefficient $R^2=0.97$ and $p=1.14~10^{-7}$. Since the effects of SRO, undetected phases and defects will naturally be higher in concentrated multicomponent alloys compared to binary alloys, we see a greater deviation at higher $\mathrm{S_{conf(per atom)}}$ values. To make our predictions more accurate, we use a more localized fit excluding high-entropy alloys for predicting $\mathrm{\rho}$ for $\mathrm{Nb_{90}Mo_{10}}$, $\mathrm{Nb_{95}Mo_{5}}$ and $\mathrm{Ta_{90}W_{10}}$ alloys as shown in Fig. \ref{figHEA_rho_residual}. The localized equation is given by $\mathrm{\rho_{r}^{alloy} =}$ $\mathrm{31.06(S_{conf(per atom)}^{1.96}) + 0.944}$ .

\subsection{Electronic thermal conductivity $\mathrm{\kappa_{e}}$}
The BoltzTrap2 package was used to solve the Boltzmann transport equation (BTE) and compute the electrical resistivity as a function of temperature using the relaxation time approximation. The values of resistivity computed in the previous section are then used to extract the values of the relaxation time through the following expression:
\begin{equation}
    \label{boltztrap2}
    \tau = \frac{(\rho \tau(T))_{BoltzTrap2}}{\rho _{model}}
\end{equation}
This relaxation time is then used to obtain the electronic part of the thermal conductivity $\mathrm{\kappa_{e}}$ from the output of BoltzTrap2.

\subsection{Lattice thermal conductivity $\mathrm{\kappa_{e}}$ and specific heat capacity $\mathrm{C_{p}}$}
We use the VASP+Phonopy package \cite{PhysRevB.47.558,phonopy-phono3py-JPCM} to compute the force constants and then use ShengBTE to solve the BTE under the RTA approximation to obtain the lattice contribution to the thermal conductivity $\mathrm{\kappa_{l}}$ and the specific heat $\mathrm{C_{p}}$. The results are then combined with $\mathrm{\kappa_{e}}$ and the total thermal conductivity is obtained as a function of temperature.

\section{Results}

\subsection{Resistivity $\mathrm{\rho}$}

\begin{figure}[H]
    \centering
    \includegraphics[width=1.05\textwidth]{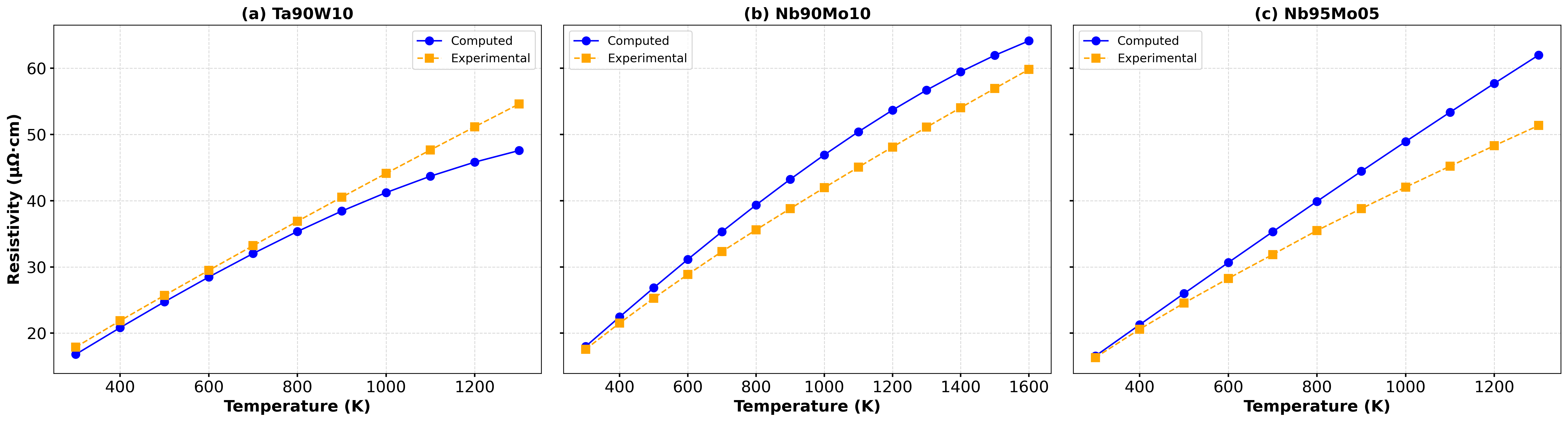} 
    \caption{Predicted resistivity and experimental resistivity for $\mathrm{Nb_{90}Mo_{10}}$, $\mathrm{Nb_{95}Mo_{05}}$ and $\mathrm{Ta_{90}W_{10}}$.}
    \label{predicted_resistivity}
\end{figure}

The resistivity of $\mathrm{Nb_{90}Mo_{10}}$, $\mathrm{Nb_{95}Mo_{05}}$ and $\mathrm{Ta_{90}W_{10}}$ has been computed using our framework and compared with experimental values. The results for room temperature resistivity predictions have been tabulated in Table \ref{tab:alloys_resistivity_300K}. The values of $A$ are obtained using DFT methods and the results are shown in Table \ref{tab:alloy_dos_A}. Finally, the values for the resistivity in the temperature range 300-1300 K are computed using Eq. \ref{eq3}. We observe an excellent statistical correlation with the experimental values as seen in Table \ref{tab:alloys_resistivity_300K} and illustrated in Fig. \ref{predicted_resistivity}. \par

\begin{figure}[htbp]
    \centering
    \includegraphics[width=1.2\textwidth]
    {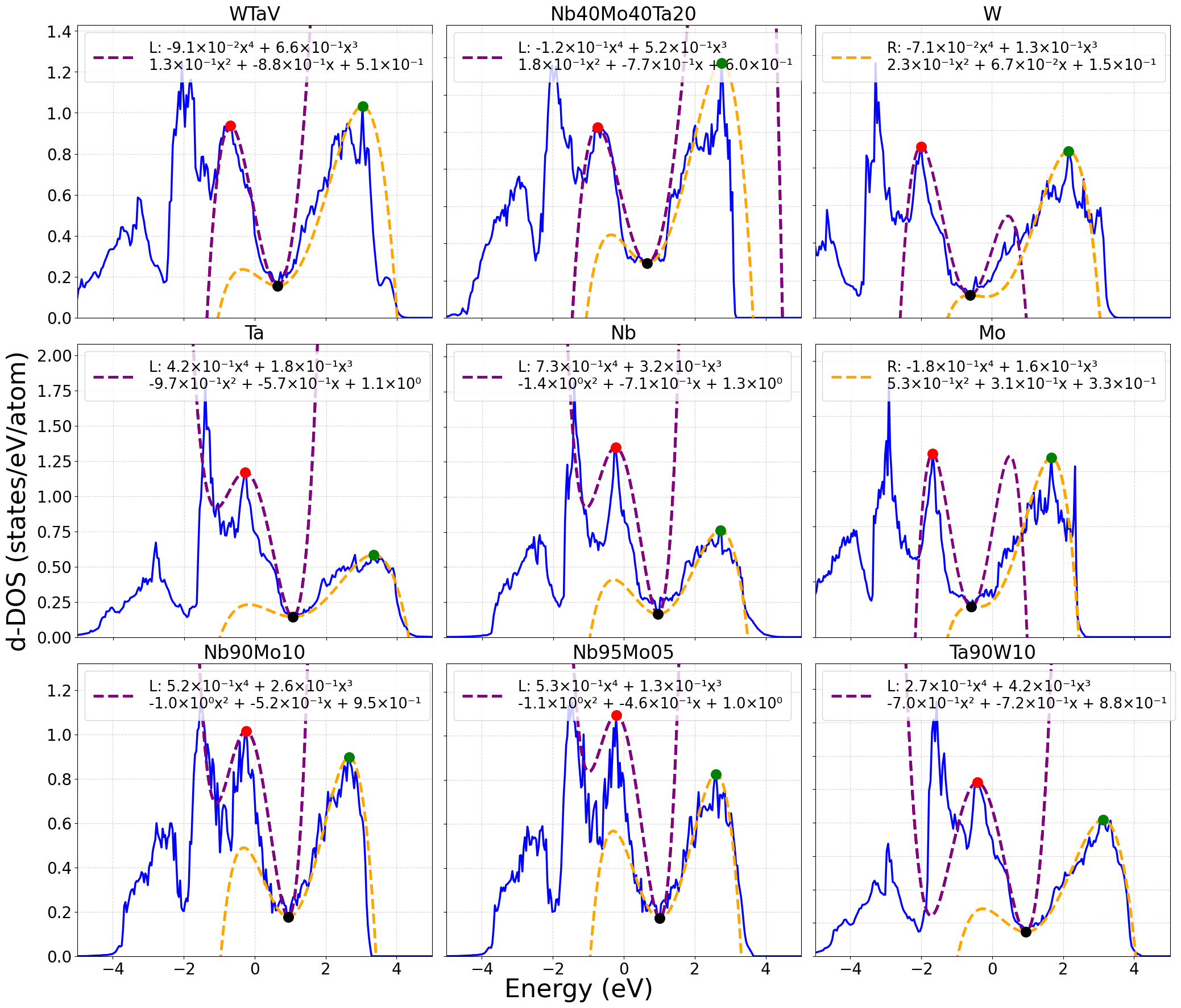}
    \caption{Mathematical smoothening of DOS plots. The equation in the legend shows the curve used to compute A, depending on wether $\mathrm{\epsilon_{F}}$ to the Left(L) or Right(R) of the trough point.}
    \label{figA_calc}
\end{figure}

\begin{table}[h]
\centering
\caption{Predicted resistivity of alloys at room temperature (300K)}
\label{tab:alloys_resistivity_300K}
\begin{tabular}{l c c c}
\hline
\textbf{Alloy} & $\mathbf{\rho}$\textbf{(predicted)}($\mathrm{\mu \Omega cm}$) & $\mathbf{\rho}$\textbf{(experimental)}($\mathrm{\mu \Omega cm}$) & $\mathbf{\Delta}$\% \\
\hline
$\mathrm{Nb_{95}Mo_{05}}$ & 16.45 & 16.31 & 0.85 \\
$\mathrm{Nb_{90}Mo_{10}}$  & 18.13 & 17.56 & 3.24 \\
$\mathrm{Ta_{90}W_{10}}$ & 16.99 & 17.89 & -5.1 \\
$\mathrm{Nb_{97}Mo_{03}}$  & 16.14 & 16.17 & -0.2 \\
\hline
\end{tabular}
\end{table}

\begin{table}[h]
\centering
\caption{Alloy $A$ values along with error analysis for resistivity trend predictions from figure \ref{predicted_resistivity}}
\label{tab:alloy_dos_A}
\begin{tabular}{l c c c c c}
\hline
\textbf{Alloy} & \textbf{$A_{model} (K^{-2})\times10^{-8}$} & \textbf{$A (K^{-2})\times10^{-8}$} & \textbf{RMSE ($\bm{\mu\Omega\,\mathrm{cm}}$)} & \textbf{MAPE ($\bm{\%}$)} & $\mathbf{R^2}$ \\
\hline
$\mathrm{Nb_{95}Mo_{05}}$ & 4.43 & 2.2 & 5.92 & 11.9 & 0.71 \\
$\mathrm{Nb_{90}Mo_{10}}$  & 5.48 & 7.3 & 4.14 & 8.87 & 0.89 \\
$\mathrm{Ta_{90}W_{10}}$ & 6.55 & 12.5 & 3.22 & 6.29 & 0.92 \\
\hline
\end{tabular}
\end{table}

\begin{table}[h]
\centering
\caption{Experimental values of $\mathrm{\rho_{r}^{alloy}}$ for HEAs}
\label{tab:error_table_residual}
\begin{tabular}{l c c c c}
\hline
\textbf{HEA} & $\mathbf{S_{conf}/k_{b}}$ ($\times$10$^{-2}$)&$\mathbf{\rho_{i}}$($\bm{\mu\Omega\,\mathrm{cm}}$)& $\mathbf{\rho_{r}^{alloy}}$($\bm{\mu\Omega\,\mathrm{cm}}$) & $\mathbf{\rho_{exp}}$($\bm{\mu\Omega\,\mathrm{cm}}$)\\

\hline
\label{tab:electrical_resistivity_prediction}
$\mathrm{Nb_{95}Mo_{05}}$&19.85 & 14.99& 1.54 & 16.31\\
$\mathrm{Nb_{90}Mo_{10}}$&32.51 & 13.99& 2.26 &   17.56\\
$\mathrm{Ta_{90}W_{10}}$&32.51 & 12.19& 1.53 &  17.89 \\
$\mathrm{W_{0.4}FeCrCo}$&133.16 & 9.02& 1.54 & 160$\pm$1.5  \\
$\mathrm{W_{2.1}FeCrCo}$&132.37 & 7.68& 1.54 &  170$\pm$1.05 \\
$\mathrm{AlCoCrFeNi}$&160.94 & 7.66& 1.54 &  220.82 \\
$\mathrm{Al2CoCrFeNi}$&156.07 & 6.81& 1.54 &  211.29 \\
$\mathrm{Al1.25CoCrFeNi}$&160.51 & 7.41& 1.54 &  167.41 \\
$\mathrm{Al1.5CoCrFeNi}$&159.42 & 7.20& 1.54 &  153.41 \\
\hline
\end{tabular}
\end{table}

\subsection{Thermal conductivity $\mathrm{\kappa}$ and specific heat $\mathrm{C_{p}}$}

We use BoltzTrap2 package to solve the BTE under a fixed relaxation time approximation for the same $\mathrm{Ta_{90}W_{10}}$ alloy. This gives us the electronic properties in the form of $\mathrm{\rho / \tau}$ and $\mathrm{\kappa_{e} / \tau}$ in the temperature range of 300 - 1300 K. We now use the absolute values of $\mathrm{\rho}$ computed previously and combine them with the BTE solution to extract the values of the electron relaxation time $\mathrm{\tau}$ as a function of temperature. This is then used to get the absolute values from the fixed relaxation time approximation solution. The lattice properties such as $\mathrm{\kappa_{l}}$ and $\mathrm{C_{p}}$ were calculated using the ShengBTE package for numerically solving the BTE for phonons. The results are tabulated in the Table \ref{tab:alloys_thermal_conductivity}.

\begin{table}[h]
\centering
\caption{Predicted thermal conductivity of $\mathrm{Ta_{90}W_{10}}$}
\label{tab:alloys_thermal_conductivity}
\begin{tabular}{c c c c c}
\hline
\textbf{Temperature} & $\mathbf{\kappa_{l}}$($\mathrm{W/ mK}$)  & $\mathbf{\kappa_{e}}$($\mathrm{W/ mK}$) & $\mathbf{\kappa}$($\mathrm{W/ mK}$) & $\mathbf{\kappa^{exp}}$($\mathrm{W/ mK}$) \\
\hline
300  & 0.57 & 39.56 & 40.13 & 46.32 \\
320  & 0.43 & 40.41 & 40.84 & 47.02 \\
340  & 0.51 & 41.20 & 41.71 & 47.61 \\
360  & 0.58 & 41.90 & 42.48 & 48.10 \\
380  & 0.46 & 42.96 & 43.42 & 48.50 \\
400  & 0.43 & 43.20 & 43.63 & 48.75 \\
\hline
\multicolumn{5}{c}{{RMSE = 5.69 ($\mathrm{W/ mK}$), MAPE = 11.9267\%, r (Pearson) = 0.99}} \\
\hline

\hline
\end{tabular}
\end{table}

\subsection{Check for resistivity saturation and validity of Wiedemann-Franz relation}

\begin{figure}[htbp]
\makebox[\textwidth][c]{%
\begin{minipage}{\textwidth}
\centering

\begin{subfigure}{0.5\textwidth}
    \centering
    \includegraphics[width=\linewidth]{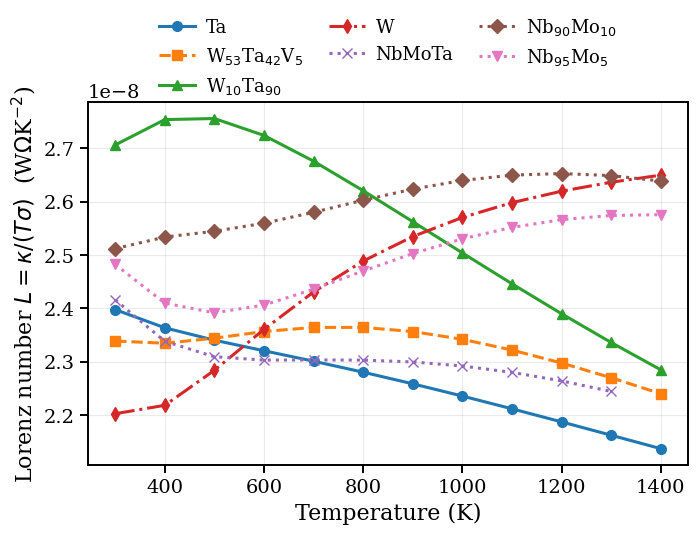}
    \caption{Wiedemann--Franz test}
    \label{figWF_test}
\end{subfigure}\hfill
\begin{subfigure}{0.5\textwidth}
    \centering
    \includegraphics[width=\linewidth]{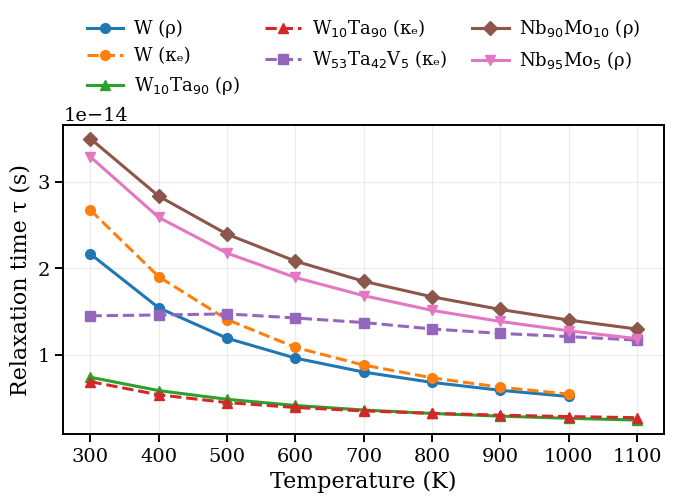}
    \caption{Relaxation times}
    \label{figrelaxation_times}
\end{subfigure}\hfill
\begin{subfigure}{0.6\textwidth}
    \centering
    \includegraphics[width=\linewidth]{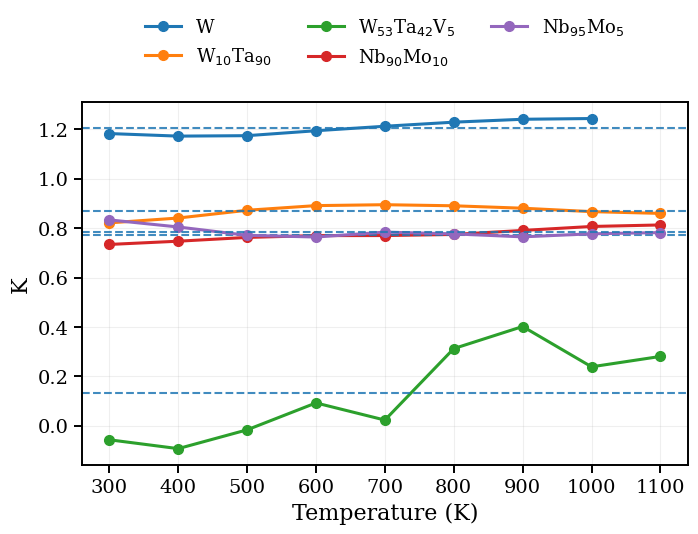}
    \caption{K parameter}
    \label{figK parameter}
\end{subfigure}

\end{minipage}}
\caption{Transport analysis: (a) Wiedemann--Franz test, (b) relaxation time behavior, and (c) local transport exponent.}
\label{figtransport_summary}
\end{figure}

The electronic relaxation times for all the alloys in this work were computed using experimental data to check the Ioffe-Regel limit with the results shown in Fig. \ref{figrelaxation_times}. The $\mathrm{\kappa_{e}}$ and $\mathrm{\rho}$ in the legend specifies wether the thermal conductivity or electrical resistivity from experiments were used in the calculation. This data is used to compute the K parameter (see Eq. \ref{eq:Kparam}), and the results show that most alloys are far from reaching the saturation resistivity. However, the $\mathrm{W_{53}Ta_{42}V_{5}}$ system has an average K value close to 0 as computed using experimental data on thermal conductivity (see Fig. \ref{figK parameter}). This indicates that the alloy saturates at temperature lower than 300K and this effect has been taken into account in the following section about alloy design. The validity of Wiedemann Franz law has also been checked for all the alloys and metals in this study. Figure \ref{figWF_test} shows that all the alloys including $\mathrm{W_{53}Ta_{42}V_{5}}$ and $\mathrm{Nb_{40}Mo_{40}Ta_{20}}$ obey the Wiedemann-Franz relation throughout the working temperature regime.

\begin{table}[h]
\centering
\caption{K values for alloys}
\label{tab:K_table}
\begin{tabular}{c c}
\hline

\textbf{Alloy} & K \\

\hline
W& 1.206  \\
$\mathrm{Nb_{90}Mo_{10}}$& 0.774  \\
$\mathrm{Ta_{90}W_{10}}$& 0.869 \\
$\mathrm{Nb_{95}Mo_{05}}$& 0.784  \\
$\mathrm{W_{53}Ta_{42}V_{05}}$& 0.132  \\
\hline
\end{tabular}
\end{table}

\subsection{Experimental validation on the $\mathrm{W_{53}Ta_{42}V_{5}}$ system}

In this section, we use our methodology to predict the properties of a model $\mathrm{W_{53}Ta_{42}V_{5}}$ alloy which was recently manufactured by { ROBIN } et al. The A value for this material is computed to be much higher than group V metals ($\mathrm{5.9 \times 10^{-7}}$) which means that the scattering to d-orbitals is very less likely, resulting in a suppression in the resistivity increase rate with temperature. Since the Wiedemann-Franz relation is also expected to be fully valid \ref{figWF_test}, the result of a slow increase in $\mathrm{\rho}$ will result in an increase in $\mathrm{\kappa_{e}}$ and hence the total $\mathrm{\kappa}$, which has been observed in experiments {Robin et al}. But, we know from our results that this alloy has already reached the Ioffe-Regel limit (\ref{tab:K_table}), we do not use the resistivity framework described above like for the other alloys, which predicts $\mathrm{\rho = 33 ~\mu \Omega cm}$ at 300 K (much higher than the saturated value). Instead, we have experimental values for thermal conductivity and specific heat over the temperature range from room temperature to 1300 K for this alloy. We use the experimental $\mathrm{\kappa}$ along with the computed $\mathrm{\kappa_{l}}$ to obtain $\mathrm{\kappa_{e}}$. We then make use of the Wiedemann-Franz law to back compute our resistivity values. The results are listed in Table \ref{tab:WTaV_thermal_conductivity}.

\begin{table}[h]
\centering
\caption{Predicted properties of $\mathrm{W_{53}Ta_{42}V_{5}}$ ($\mathrm{A=6.07\times 10^{-7}}K^{-2})$}
\label{tab:WTaV_thermal_conductivity}
\begin{tabular}{c c c c c}
\hline
\textbf{Temperature} & $\mathbf{\kappa_{l}}$($\mathrm{W/ mK}$)  & $\mathbf{\kappa^{exp}}$($\mathrm{W/ mK}$) & $\mathbf{\kappa_{e}}$($\mathrm{W/ mK}$) & $\mathbf{\rho}$($\bm{\mu\Omega\,\mathrm{cm}}$) \\
\hline
300 & 16.01 & 48.48 & 37.38 & 20.73 \\
400 & 11.82 & 53.47 & 49.79 &  20.52\\
500 & 9.39 & 62.96 & 62.71 &  20.08\\
600 & 7.80 & 67.92 & 72.93 &  20.41\\
700 & 6.67 & 74.76 & 81.99 &  20.82\\
800 & 5.83 & 81.24 & 88.51 &  20.70\\
900 & 5.18 & 86.55 & 95.18 &  22.33\\
1000 & 4.66 & 92.56 & 101.91 &  22.80\\
1100 & 4.23 & 98.14 & 107.39 &  23.43\\
1200 & 3.88 & 103.93 & 113.56 &  23.80\\
1300 & 3.58 & 110.29 & 119.07 &  24.22\\
\hline
\end{tabular}
\end{table}

As established in the previous section, our predictions show that the resistivity of this alloy does not change much with temperature and has reached a saturation point. This is an interesting result, as this material behavior resembles Zero-Temperature Coefficient (ZTE) alloys \cite{Shafeie2019,Lee2019} and is highly useful for precision applications. We plan to test these predictions in the future.  

\subsection{Alloy design $\mathrm{Nb_{40}Mo_{40}Ta_{20}}$}
\begin{figure}[H]
    \centering
    \begin{subfigure}{0.32\textwidth}
        \centering
        \includegraphics[width=\linewidth]{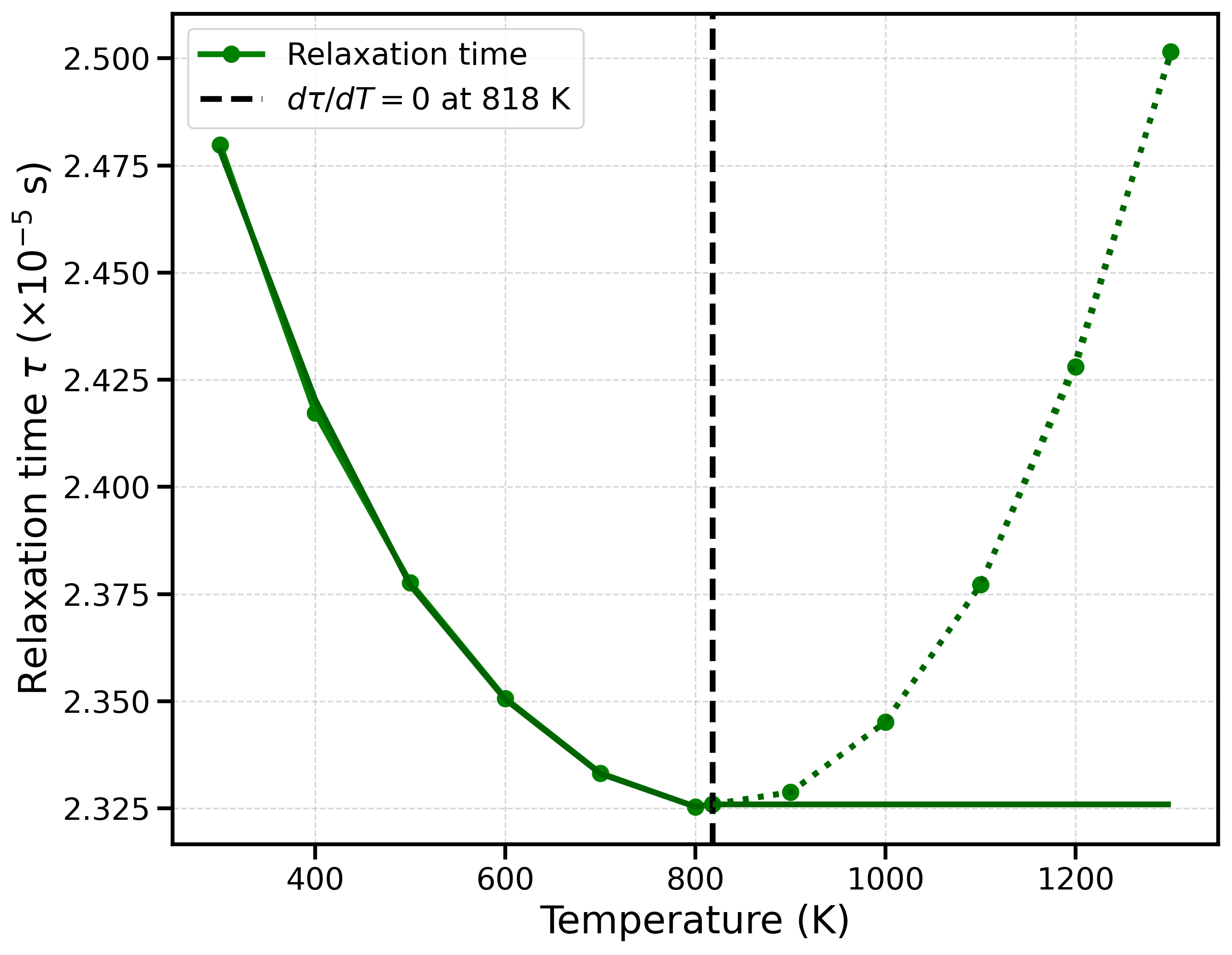}
        \caption{predicted relaxation time vs temperature}
        \label{figNbMoTa-a}
    \end{subfigure}
    \hfill
    \begin{subfigure}{0.32\textwidth}
        \centering
        \includegraphics[width=\linewidth]{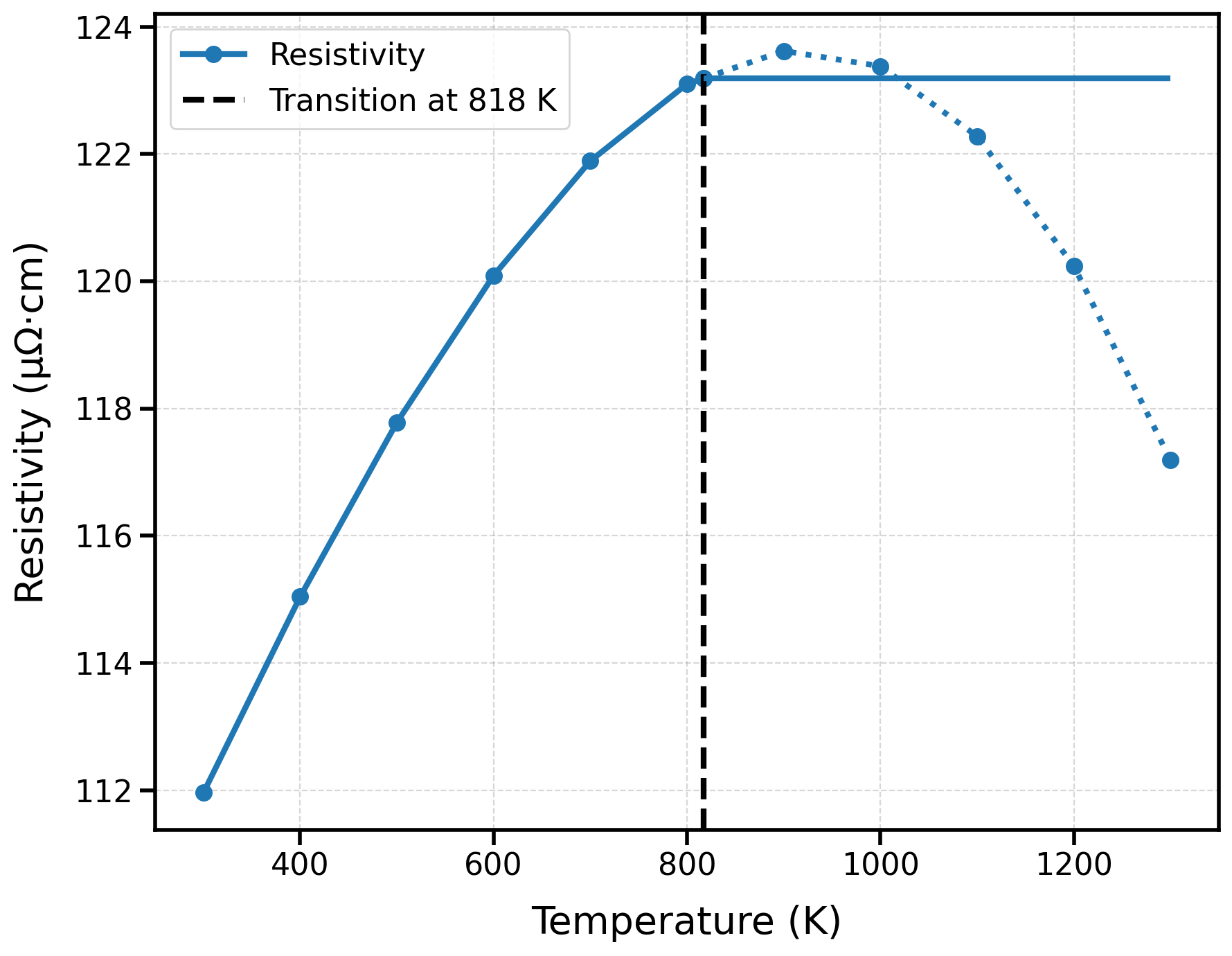}
        \caption{predicted electrical resistivity vs temperature}
        \label{figNbMoTa-b}
    \end{subfigure}
    \hfill
    \begin{subfigure}{0.32\textwidth}
        \centering
        \includegraphics[width=\linewidth]{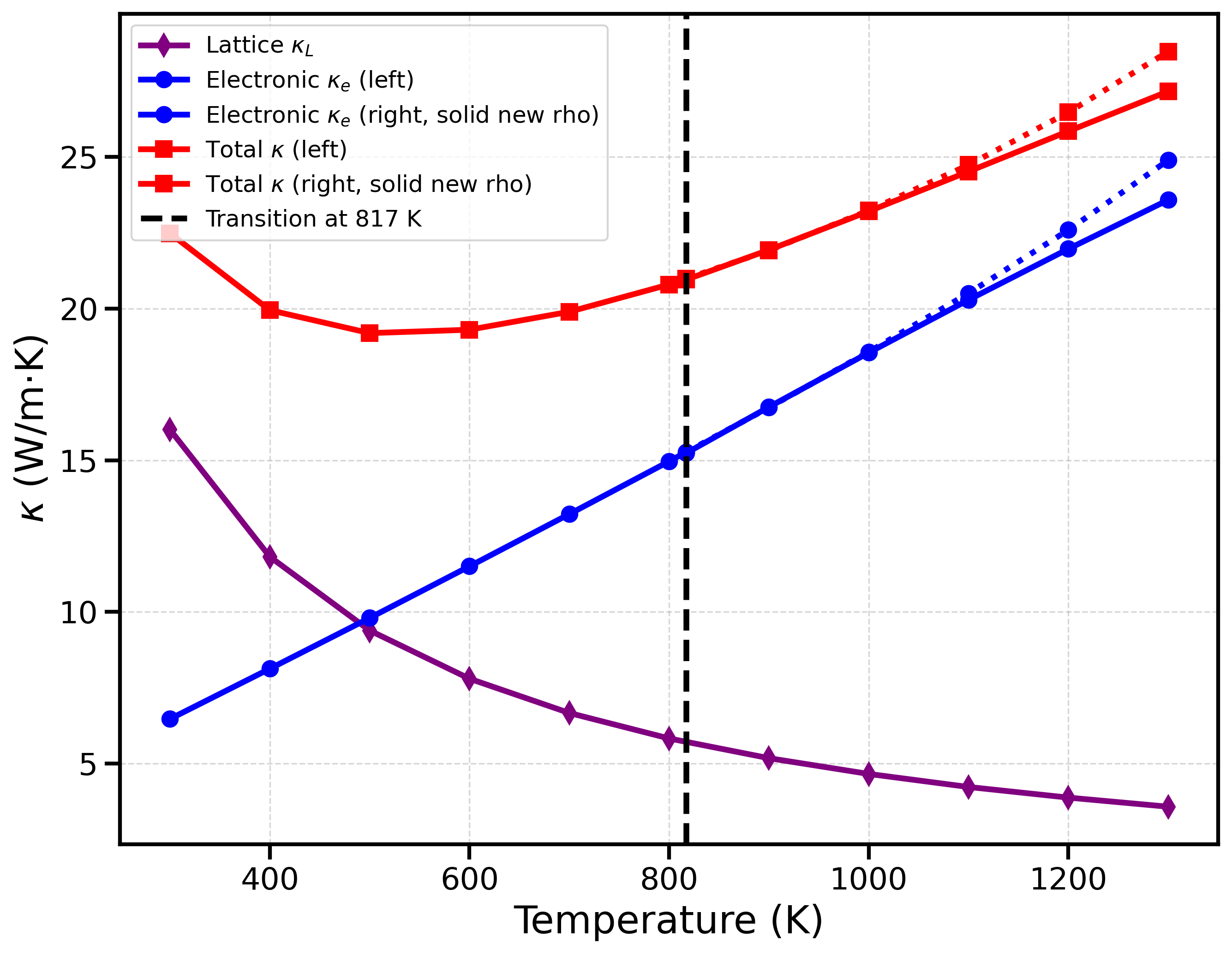}
        \caption{Predicted thermal conductivity vs temperature}
        \label{figNbMoTa-c}
    \end{subfigure}
    \caption{Prediction of thermo-electric properties of $\mathrm{Nb_{40}Mo_{40}Ta_{20}}$. The dotted line represent predictions without the s-d saturation discussed in \ref{s-d}.}

    \label{figNbMoTa}
\end{figure}

Using this model, we have predicted the thermo-electric properties of a $\mathrm{Nb_{40}Mo_{40}Ta_{20}}$ system. Since we have no data available for this alloy family, we could not check for resistivity saturation, but this alloy is assumed to have a high Ioffe-Regel limit like the binary alloys from previous sections. Our goal with this alloy was to develop a material for a heating element that operates at a temperature of 1300 K. Some key thermo-electric requirements were high resistivity for high Joule heating, constant thermal conductivity with temperature to prevent thermal stresses and a high melting point preventing thermal creep. We use our framework to predict the  thermal and electrical properties in the temperature range of 300-1300 K. The results are tabulated in Table \ref{tab:NbMoTa predicted properties}. Our predictions show that the alloy is expected to exhibit a high $\mathrm{\rho}$ and a near constant $\mathrm{\kappa}$ (20-25 W/mK) in the temperature regime of 300K-1300K. However, since the A value is quite large for this alloy ($\mathrm{3.94\times 10^{-7} K^{-2}}$), we expect the $\mathrm{\rho}$ to show a suppressed increase with temperature. We do the validity checks from \ref{checks} and as mentionde previously, we assume that the alloy does not reach the Ioffe-Regel limit in this temperature regime. The alloy is also expected to follow the Wiedemann-Franz law as seen in \ref{figWF_test}. For testing the s-d saturation (\ref{s-d}), we plot the $\mathrm{\tau}$ vs T as seen in figure \ref{figNbMoTa-a} and we see that there is an inflection point at 818 K. We treat this temperature point to be the cut-off for the $A$ parameter and turn it to zero. The predicted properties of this alloy can be found in \ref{tab:NbMoTa predicted properties} and \ref{figNbMoTa}. We are in the process of manufacturing this alloy and plan to validate our predictions in the immediate future.

\begin{table}[h]
\centering
\caption{Predicted properties of $\mathrm{Nb_{40}Mo_{40}Ta_{20}}$$(\mathrm{A=3.94\times 10^{-7} K^{-2}})$}
\label{tab:NbMoTa predicted properties}
\begin{tabular}{c c c c c c}
\hline
\textbf{Temperature (K)} & \textbf{$\kappa_{l}$ (W/mK)} & \textbf{$\kappa_{e}$ (W/mK)} & \textbf{$\kappa$ (W/mK)} & \textbf{$\rho$ ($\mu\Omega\cdot$cm)} & \textbf{C$_p$$\times10^{6}$ (J/m$^3$K)} \\
\hline
300  & 16.01 & 6.47 & 22.48 & 111.96  & $2.29$ \\
400  & 11.82 & 8.13 & 19.95 & 115.04  & $2.35$ \\
500  & 9.39  & 9.80 & 19.19 & 117.78  & $2.38$ \\
600  & 7.80  & 11.50 & 19.30 & 120.09  & $2.39$ \\
700  & 6.67  & 13.23 & 19.90 & 121.89  & $2.40$ \\
800  & 5.83  & 14.96 & 20.79 & 121.89  & $2.40$ \\
900  & 5.18  & 16.75 & 21.93 & 123.19 & $2.40$ \\
1000 & 4.66  & 18.55 & 23.21 & 123.19 & $2.40$ \\
1100 & 4.23  & 20.29 & 24.52 & 123.19 & $2.40$ \\
1200 & 3.88  & 21.97 & 25.85 & 123.19 & $2.40$ \\
1300 & 3.58  & 23.58 & 27.16 & 123.19 & $2.40$ \\
\hline
\end{tabular}
\end{table}

\vspace{0.5cm}

\section{Discussion and Conclusions}
In this work, we proposed a physics-informed framework for an efficient estimation of the thermo-electric properties of BCC concentrated alloys. This model enables us to predict the thermo-electric properties of bcc transition metals and their alloys with extremely efficient computational expense. We were able to validate the predictions for room temperature $\mathrm{\rho}$ for a few single phase alloys (tables \ref{tab:alloys_resistivity_300K} and \ref{tab:electrical_resistivity_prediction}) which we were sure that they have a single phase bcc structure and were expected to have similar Debye temperatures. Due to the lack of available experimental data in the temperature regime 300 - 1300 K for single phase BCC solid solutions, we were only able to validate our results as a function of temperature for a few dilute  bcc binary alloys. While our model shows excellent correlation with those systems, validation against more complex concentrated alloys is required to further improve the robustness and test the validity of this model. The $\mathrm{W_{53}Ta_{42}V_{5}}$ has already been synthesized with $\mathrm{\kappa}$ and $\mathrm{C_{p}}$ values available, thus we plan to experimentally measure $\mathrm{\rho}$ as a follow-up to this work. The $\mathrm{Nb_{40}Mo_{40}Ta_{20}}$ alloy is also predicted to be a viable candidate for applications at elevated temperature (upto 1300 K), but experimental validation is required to check for resistivity saturation. We plan to synthesize and characterize this alloy in the near future to further test the viability of our approach. If successfully validated, this novel approach can potentially help expedite the development of concentrated alloys with specific thermo-electric properties. \par

As mentioned before, we neglect the effects of thermal expansion on the thermo-electric properties. Although mathematically, since it is linearly related, the effects of thermal expansion is expected to be small as compared to the s-d scattering for transition metals, it is still significant and should be added for further accuracy. Also, we treat the value of A to be independent of the temperature which is not true experimentally. Our model shows that it is still a good approximation and captures trends really well in the temperature regime. Finally, we also assume a constant value of $\mathrm{\Theta_{D}}$ for all metals and that it is independent of temperature. For maximum accuracy, the $\mathrm{\Theta_{D}}$ should be computed for all the alloys and also as a function of T.It needs to be mentioned that these approximations, although not drastically, are likely to induce some errors in the predicted values. We plan to compute these as a future extention of this work once we have the experimental data available for the NbMoTa and WTaV concentrated alloys discussed in this work.
\par

\par 
We also make some geometric approximations by smoothening the DOS plot and by treating the $A$ values as a constant in extracting the experimental $A$. This is likely to induce some error and is likely the reason our theoretical $A$ values and the experimental $A$ are not a one-to-one match but rather a linearly related (\ref{A_fit}). This is fine as the model is excellent at predicting trends and needs to maintain computational efficiency. \par

For the sake of succinctness and to avoid digression, we developed this model specifically for BCC alloys. The physical basis of this theory however, is perfectly translatable to other crystal structures as well without much change. The probability of scattering given by equation \ref{eq2} and the residual relation with the system disorder from equation \ref{eq8} would naturally change, along with modifications to equation \ref{eq5} to account for any crystallography specific terms (similar to the d-orbital scattering that is specific to BCC transition metals as mentioned in this work). Even so, more work is needed to extend this framework to other crystal structures, which is left for a futre work. \par

Finally, it is important to mention that this work does not account for the effects of short-range order (SRO). It is well known that most alloys exhibit SRO to some extent at low temperatures.  The presence of order in the system might result in lower than expected $\mathrm{\rho}$. This is key for explaining the ZTE behavior in alloys like $\mathrm{W_{53}Ta_{42}V_{5}}$, which are known to exhibit SRO up to 600 K [Robin]. If our predictions are true, the SRO is likely to cause a decrease in $\mathrm{\rho}$ proportional to $\mathrm{S_{conf}}$ of the ordered system, thus resulting in a modified Eq. \ref{eq4} of the form $\mathrm{\rho = \rho_{r} + \rho_{i} + \rho_{r}^{alloy} - XT^{X}}$, where, X is a term that quantifies the SRO for an alloy. Regardless, a theory that accounts for SRO effects is quite complex and is beyond the scope of this work, best left for a future study. This work also neglects crystal defects such as dislocations, grain boundaries and phase interfaces, all of which could play significant roles in the thermo-electric properties of concentrated alloys. 

\section{Declaration of competing interest}
The authors declare that they have no known competing financial interests or personal relationships that could have appeared to influence the work reported in this paper.

\section{Acknowledgment}
The authors acknowledge the support from the US Department of Energy, the Industrial Efficiency and
Decarbonization Office (IEDO) under grant number DE‐EE0011197.
Clemson University is acknowledged for generous allotment of compute time on the Palmetto cluster. This material is based on work supported by the National Science Foundation under Grant Nos. MRI\# 2024205, MRI\# 1725573, and CRI\# 2010270 \cite{antao2024modernizing}.

\begin{appendices}

\section{Section title of first appendix}\label{secA1}

\end{appendices}

\bibliography{sn-bibliography}


\begin{thebibliography}{31}
\ifx \bisbn   \undefined \def \bisbn  #1{ISBN #1}\fi
\ifx \binits  \undefined \def \binits#1{#1}\fi
\ifx \bauthor  \undefined \def \bauthor#1{#1}\fi
\ifx \batitle  \undefined \def \batitle#1{#1}\fi
\ifx \bjtitle  \undefined \def \bjtitle#1{#1}\fi
\ifx \bvolume  \undefined \def \bvolume#1{\textbf{#1}}\fi
\ifx \byear  \undefined \def \byear#1{#1}\fi
\ifx \bissue  \undefined \def \bissue#1{#1}\fi
\ifx \bfpage  \undefined \def \bfpage#1{#1}\fi
\ifx \blpage  \undefined \def \blpage #1{#1}\fi
\ifx \burl  \undefined \def \burl#1{\textsf{#1}}\fi
\ifx \doiurl  \undefined \def \doiurl#1{\url{https://doi.org/#1}}\fi
\ifx \betal  \undefined \def \betal{\textit{et al.}}\fi
\ifx \binstitute  \undefined \def \binstitute#1{#1}\fi
\ifx \binstitutionaled  \undefined \def \binstitutionaled#1{#1}\fi
\ifx \bctitle  \undefined \def \bctitle#1{#1}\fi
\ifx \beditor  \undefined \def \beditor#1{#1}\fi
\ifx \bpublisher  \undefined \def \bpublisher#1{#1}\fi
\ifx \bbtitle  \undefined \def \bbtitle#1{#1}\fi
\ifx \bedition  \undefined \def \bedition#1{#1}\fi
\ifx \bseriesno  \undefined \def \bseriesno#1{#1}\fi
\ifx \blocation  \undefined \def \blocation#1{#1}\fi
\ifx \bsertitle  \undefined \def \bsertitle#1{#1}\fi
\ifx \bsnm \undefined \def \bsnm#1{#1}\fi
\ifx \bsuffix \undefined \def \bsuffix#1{#1}\fi
\ifx \bparticle \undefined \def \bparticle#1{#1}\fi
\ifx \barticle \undefined \def \barticle#1{#1}\fi
\bibcommenthead
\ifx \bconfdate \undefined \def \bconfdate #1{#1}\fi
\ifx \botherref \undefined \def \botherref #1{#1}\fi
\ifx \url \undefined \def \url#1{\textsf{#1}}\fi
\ifx \bchapter \undefined \def \bchapter#1{#1}\fi
\ifx \bbook \undefined \def \bbook#1{#1}\fi
\ifx \bcomment \undefined \def \bcomment#1{#1}\fi
\ifx \oauthor \undefined \def \oauthor#1{#1}\fi
\ifx \citeauthoryear \undefined \def \citeauthoryear#1{#1}\fi
\ifx \endbibitem  \undefined \def \endbibitem {}\fi
\ifx \bconflocation  \undefined \def \bconflocation#1{#1}\fi
\ifx \arxivurl  \undefined \def \arxivurl#1{\textsf{#1}}\fi
\csname PreBibitemsHook\endcsname

\bibitem[\protect\citeauthoryear{Hu et~al.}{2021}]{HU2021116800}
\begin{barticle}
\bauthor{\bsnm{Hu}, \binits{Y.-J.}},
\bauthor{\bsnm{Sundar}, \binits{A.}},
\bauthor{\bsnm{Ogata}, \binits{S.}},
\bauthor{\bsnm{Qi}, \binits{L.}}:
\batitle{Screening of generalized stacking fault energies, surface energies and intrinsic ductile potency of refractory multicomponent alloys}.
\bjtitle{Acta Materialia}
\bvolume{210},
\bfpage{116800}
(\byear{2021})
\doiurl{10.1016/j.actamat.2021.116800}
\end{barticle}
\endbibitem

\bibitem[\protect\citeauthoryear{Korpe et~al.}{2025}]{KORPE2025102519}
\begin{botherref}
\oauthor{\bsnm{Korpe}, \binits{A.}},
\oauthor{\bsnm{El-Atwani}, \binits{O.}},
\oauthor{\bsnm{Martinez}, \binits{E.}}:
Effect of alloying on intrinsic ductility in wtacrv high entropy alloys.
Materialia,
102519
(2025)
\doiurl{10.1016/j.mtla.2025.102519}
\end{botherref}
\endbibitem

\bibitem[\protect\citeauthoryear{Mak et~al.}{2021}]{MAK2021104389}
\begin{barticle}
\bauthor{\bsnm{Mak}, \binits{E.}},
\bauthor{\bsnm{Yin}, \binits{B.}},
\bauthor{\bsnm{Curtin}, \binits{W.A.}}:
\batitle{A ductility criterion for bcc high entropy alloys}.
\bjtitle{Journal of the Mechanics and Physics of Solids}
\bvolume{152},
\bfpage{104389}
(\byear{2021})
\doiurl{10.1016/j.jmps.2021.104389}
\end{barticle}
\endbibitem

\bibitem[\protect\citeauthoryear{Singh et~al.}{2023}]{SINGH2023119104}
\begin{barticle}
\bauthor{\bsnm{Singh}, \binits{P.}},
\bauthor{\bsnm{Vela}, \binits{B.}},
\bauthor{\bsnm{Ouyang}, \binits{G.}},
\bauthor{\bsnm{Argibay}, \binits{N.}},
\bauthor{\bsnm{Cui}, \binits{J.}},
\bauthor{\bsnm{Arroyave}, \binits{R.}},
\bauthor{\bsnm{Johnson}, \binits{D.D.}}:
\batitle{A ductility metric for refractory-based multi-principal-element alloys}.
\bjtitle{Acta Materialia}
\bvolume{257},
\bfpage{119104}
(\byear{2023})
\doiurl{10.1016/j.actamat.2023.119104}
\end{barticle}
\endbibitem

\bibitem[\protect\citeauthoryear{Li et~al.}{2020}]{LI2020174}
\begin{barticle}
\bauthor{\bsnm{Li}, \binits{X.}},
\bauthor{\bsnm{Li}, \binits{W.}},
\bauthor{\bsnm{Irving}, \binits{D.L.}},
\bauthor{\bsnm{Varga}, \binits{L.K.}},
\bauthor{\bsnm{Vitos}, \binits{L.}},
\bauthor{\bsnm{Schönecker}, \binits{S.}}:
\batitle{Ductile and brittle crack-tip response in equimolar refractory high-entropy alloys}.
\bjtitle{Acta Materialia}
\bvolume{189},
\bfpage{174}--\blpage{187}
(\byear{2020})
\doiurl{10.1016/j.actamat.2020.03.004}
\end{barticle}
\endbibitem

\bibitem[\protect\citeauthoryear{Madsen et~al.}{2018}]{BoltzTraP2}
\begin{barticle}
\bauthor{\bsnm{Madsen}, \binits{G.K.H.}},
\bauthor{\bsnm{Carrete}, \binits{J.}},
\bauthor{\bsnm{Verstraete}, \binits{M.J.}}:
\batitle{{BoltzTraP2}, a program for interpolating band structures and calculating semi-classical transport coefficients}.
\bjtitle{Comput. Phys. Commun.}
\bvolume{231},
\bfpage{140}--\blpage{145}
(\byear{2018})
\doiurl{10.1016/j.cpc.2018.05.010}
\end{barticle}
\endbibitem

\bibitem[\protect\citeauthoryear{Noffsinger et~al.}{2010}]{NOFFSINGER20102140}
\begin{barticle}
\bauthor{\bsnm{Noffsinger}, \binits{J.}},
\bauthor{\bsnm{Giustino}, \binits{F.}},
\bauthor{\bsnm{Malone}, \binits{B.D.}},
\bauthor{\bsnm{Park}, \binits{C.-H.}},
\bauthor{\bsnm{Louie}, \binits{S.G.}},
\bauthor{\bsnm{Cohen}, \binits{M.L.}}:
\batitle{Epw: A program for calculating the electron–phonon coupling using maximally localized wannier functions}.
\bjtitle{Computer Physics Communications}
\bvolume{181}(\bissue{12}),
\bfpage{2140}--\blpage{2148}
(\byear{2010})
\doiurl{10.1016/j.cpc.2010.08.027}
\end{barticle}
\endbibitem

\bibitem[\protect\citeauthoryear{Ziman}{1967}]{ziman1967electrons}
\begin{bbook}
\bauthor{\bsnm{Ziman}, \binits{J.M.}}:
\bbtitle{Electrons and Phonons: The Theory of Transport Phenomena in Solids}.
\bsertitle{International series of monographs on physics}.
\bpublisher{Clarendon Press}, \blocation{???}
(\byear{1967}).
\burl{https://books.google.com/books?id=wFI5wAEACAAJ}
\end{bbook}
\endbibitem

\bibitem[\protect\citeauthoryear{Chiu}{1976}]{PhysRevB.13.1507}
\begin{barticle}
\bauthor{\bsnm{Chiu}, \binits{J.C.H.}}:
\batitle{Deviations from linear temperature dependence of the electrical resistivity of v-cr and ta-w alloys}.
\bjtitle{Phys. Rev. B}
\bvolume{13},
\bfpage{1507}--\blpage{1514}
(\byear{1976})
\doiurl{10.1103/PhysRevB.13.1507}
\end{barticle}
\endbibitem

\bibitem[\protect\citeauthoryear{Desai et~al.}{1984}]{Desai10.1063/1.555723}
\begin{barticle}
\bauthor{\bsnm{Desai}, \binits{P.D.}},
\bauthor{\bsnm{Chu}, \binits{T.K.}},
\bauthor{\bsnm{James}, \binits{H.M.}},
\bauthor{\bsnm{Ho}, \binits{C.Y.}}:
\batitle{Electrical resistivity of selected elements}.
\bjtitle{Journal of Physical and Chemical Reference Data}
\bvolume{13}(\bissue{4}),
\bfpage{1069}--\blpage{1096}
(\byear{1984})
\doiurl{10.1063/1.555723}
{\href{https://arxiv.org/abs/https://pubs.aip.org/aip/jpr/article-pdf/13/4/1069/9766234/1069\_1\_online.pdf}{{https://pubs.aip.org/aip/jpr/article-pdf/13/4/1069/9766234/1069\_1\_online.pdf}}}
\end{barticle}
\endbibitem

\bibitem[\protect\citeauthoryear{P.}{1937}]{P1937}
\begin{barticle}
\bauthor{\bsnm{P.}, \binits{R.}}:
\batitle{The theory of the properties of metals and alloys}.
\bjtitle{Nature}
\bvolume{139}(\bissue{3513}),
\bfpage{348}--\blpage{349}
(\byear{1937})
\doiurl{10.1038/139348a0}
\end{barticle}
\endbibitem

\bibitem[\protect\citeauthoryear{Yamashita and Asano}{1974}]{10.1143/PTP.51.317b}
\begin{barticle}
\bauthor{\bsnm{Yamashita}, \binits{J.}},
\bauthor{\bsnm{Asano}, \binits{S.}}:
\batitle{Electrical resistivity of transition metals. i}.
\bjtitle{Progress of Theoretical Physics}
\bvolume{51}(\bissue{2}),
\bfpage{317}--\blpage{326}
(\byear{1974})
\doiurl{10.1143/PTP.51.317b}
{\href{https://arxiv.org/abs/https://academic.oup.com/ptp/article-pdf/51/2/317/5302245/51-2-317.pdf}{{https://academic.oup.com/ptp/article-pdf/51/2/317/5302245/51-2-317.pdf}}}
\end{barticle}
\endbibitem

\bibitem[\protect\citeauthoryear{Kao et~al.}{2011}]{KAO20111607}
\begin{barticle}
\bauthor{\bsnm{Kao}, \binits{Y.-F.}},
\bauthor{\bsnm{Chen}, \binits{S.-K.}},
\bauthor{\bsnm{Chen}, \binits{T.-J.}},
\bauthor{\bsnm{Chu}, \binits{P.-C.}},
\bauthor{\bsnm{Yeh}, \binits{J.-W.}},
\bauthor{\bsnm{Lin}, \binits{S.-J.}}:
\batitle{Electrical, magnetic, and hall properties of alxcocrfeni high-entropy alloys}.
\bjtitle{Journal of Alloys and Compounds}
\bvolume{509}(\bissue{5}),
\bfpage{1607}--\blpage{1614}
(\byear{2011})
\doiurl{10.1016/j.jallcom.2010.10.210}
\end{barticle}
\endbibitem

\bibitem[\protect\citeauthoryear{Wang et~al.}{2023}]{article}
\begin{barticle}
\bauthor{\bsnm{Wang}, \binits{H.}},
\bauthor{\bsnm{Zhang}, \binits{H.}},
\bauthor{\bsnm{Liu}, \binits{M.}},
\bauthor{\bsnm{Liu}, \binits{J.}},
\bauthor{\bsnm{Yan}, \binits{Z.}},
\bauthor{\bsnm{Zhang}, \binits{C.}},
\bauthor{\bsnm{Li}, \binits{Y.}},
\bauthor{\bsnm{Feng}, \binits{J.}}:
\batitle{The effect of w content on the microstructure, mechanics and electrical performance of an fecrco alloy}.
\bjtitle{Materials}
\bvolume{16},
\bfpage{4319}
(\byear{2023})
\doiurl{10.3390/ma16124319}
\end{barticle}
\endbibitem

\bibitem[\protect\citeauthoryear{Yeh}{2006}]{yeh_article}
\begin{barticle}
\bauthor{\bsnm{Yeh}, \binits{J.-W.}}:
\batitle{Recent progress in high-entropy alloys}.
\bjtitle{European Journal of Control - EUR J CONTROL}
\bvolume{31},
\bfpage{633}--\blpage{648}
(\byear{2006})
\doiurl{10.3166/acsm.31.633-648}
\end{barticle}
\endbibitem

\bibitem[\protect\citeauthoryear{Li et~al.}{2014}]{ShengBTE_2014}
\begin{barticle}
\bauthor{\bsnm{Li}, \binits{W.}},
\bauthor{\bsnm{Carrete}, \binits{J.}},
\bauthor{\bsnm{Katcho}, \binits{N.A.}},
\bauthor{\bsnm{Mingo}, \binits{N.}}:
\batitle{{ShengBTE:} a solver of the {B}oltzmann transport equation for phonons}.
\bjtitle{Comp. Phys. Commun.}
\bvolume{185},
\bfpage{1747}--\blpage{1758}
(\byear{2014})
\doiurl{10.1016/j.cpc.2014.02.015}
\end{barticle}
\endbibitem

\bibitem[\protect\citeauthoryear{Gunnarsson et~al.}{2003}]{loffe_regel_experimental_RevModPhys.75.1085}
\begin{barticle}
\bauthor{\bsnm{Gunnarsson}, \binits{O.}},
\bauthor{\bsnm{Calandra}, \binits{M.}},
\bauthor{\bsnm{Han}, \binits{J.E.}}:
\batitle{Colloquium: Saturation of electrical resistivity}.
\bjtitle{Rev. Mod. Phys.}
\bvolume{75},
\bfpage{1085}--\blpage{1099}
(\byear{2003})
\doiurl{10.1103/RevModPhys.75.1085}
\end{barticle}
\endbibitem

\bibitem[\protect\citeauthoryear{Gurvitch}{1981}]{loffe-regel-thory-1-PhysRevB.24.7404}
\begin{barticle}
\bauthor{\bsnm{Gurvitch}, \binits{M.}}:
\batitle{Ioffe-regel criterion and resistivity of metals}.
\bjtitle{Phys. Rev. B}
\bvolume{24},
\bfpage{7404}--\blpage{7407}
(\byear{1981})
\doiurl{10.1103/PhysRevB.24.7404}
\end{barticle}
\endbibitem

\bibitem[\protect\citeauthoryear{Tzeng et~al.}{1992}]{loffe-regel-alloy-1-1992674}
\begin{barticle}
\bauthor{\bsnm{Tzeng}, \binits{S.J.}},
\bauthor{\bsnm{Lin}, \binits{J.J.}},
\bauthor{\bsnm{Yao}, \binits{Y.D.}},
\bauthor{\bsnm{Chen}, \binits{Y.Y.}}:
\batitle{Electrical resistivity of ti<sub>0.862</sub>al<sub>0.102</sub>v<sub>0.036</sub> alloy between 4 and 1000 k}.
\bjtitle{Journal of the Physical Society of Japan}
\bvolume{61}(\bissue{2}),
\bfpage{674}--\blpage{678}
(\byear{1992})
\doiurl{10.1143/JPSJ.61.674}
\end{barticle}
\endbibitem

\bibitem[\protect\citeauthoryear{Sundqvist}{2021}]{loffe-regel-alloy-2-2SUNDQVIST2021127291}
\begin{barticle}
\bauthor{\bsnm{Sundqvist}, \binits{B.}}:
\batitle{Correlation between weak localization effects and resistivity saturation in dilute ti-al alloys}.
\bjtitle{Physics Letters A}
\bvolume{400},
\bfpage{127291}
(\byear{2021})
\doiurl{10.1016/j.physleta.2021.127291}
\end{barticle}
\endbibitem

\bibitem[\protect\citeauthoryear{Hirel}{2015}]{atomsk_HIREL2015212}
\begin{barticle}
\bauthor{\bsnm{Hirel}, \binits{P.}}:
\batitle{Atomsk: A tool for manipulating and converting atomic data files}.
\bjtitle{Computer Physics Communications}
\bvolume{197},
\bfpage{212}--\blpage{219}
(\byear{2015})
\doiurl{10.1016/j.cpc.2015.07.012}
\end{barticle}
\endbibitem

\bibitem[\protect\citeauthoryear{Kresse and Furthm\"uller}{1996}]{vasp_PhysRevB.54.11169}
\begin{barticle}
\bauthor{\bsnm{Kresse}, \binits{G.}},
\bauthor{\bsnm{Furthm\"uller}, \binits{J.}}:
\batitle{Efficient iterative schemes for ab initio total-energy calculations using a plane-wave basis set}.
\bjtitle{Phys. Rev. B}
\bvolume{54},
\bfpage{11169}--\blpage{11186}
(\byear{1996})
\doiurl{10.1103/PhysRevB.54.11169}
\end{barticle}
\endbibitem

\bibitem[\protect\citeauthoryear{Wang et~al.}{2021}]{VASPKIT}
\begin{barticle}
\bauthor{\bsnm{Wang}, \binits{V.}},
\bauthor{\bsnm{Xu}, \binits{N.}},
\bauthor{\bsnm{Liu}, \binits{J.-C.}},
\bauthor{\bsnm{Tang}, \binits{G.}},
\bauthor{\bsnm{Geng}, \binits{W.-T.}}:
\batitle{Vaspkit: A user-friendly interface facilitating high-throughput computing and analysis using vasp code}.
\bjtitle{Computer Physics Communications}
\bvolume{267},
\bfpage{108033}
(\byear{2021})
\doiurl{10.1016/j.cpc.2021.108033}
\end{barticle}
\endbibitem

\bibitem[\protect\citeauthoryear{Togo et~al.}{2023}]{phonopy-phono3py-JPCM}
\begin{barticle}
\bauthor{\bsnm{Togo}, \binits{A.}},
\bauthor{\bsnm{Chaput}, \binits{L.}},
\bauthor{\bsnm{Tadano}, \binits{T.}},
\bauthor{\bsnm{Tanaka}, \binits{I.}}:
\batitle{Implementation strategies in phonopy and phono3py}.
\bjtitle{J. Phys. Condens. Matter}
\bvolume{35}(\bissue{35}),
\bfpage{353001}
(\byear{2023})
\doiurl{10.1088/1361-648X/acd831}
\end{barticle}
\endbibitem

\bibitem[\protect\citeauthoryear{Moore et~al.}{1980}]{Moore1980}
\begin{botherref}
\oauthor{\bsnm{Moore}, \binits{J.P.}},
\oauthor{\bsnm{Graves}, \binits{R.S.}},
\oauthor{\bsnm{Williams}, \binits{R.K.}}:
Thermal transport properties of niobium and some niobium base alloys from 80 to 1600\textdegree k
(1980).
Accessed: 2025-08-26; Crediting UNT Libraries Government Documents Department
\end{botherref}
\endbibitem

\bibitem[\protect\citeauthoryear{Taylor et~al.}{1971}]{TAYLOR1971369}
\begin{barticle}
\bauthor{\bsnm{Taylor}, \binits{R.E.}},
\bauthor{\bsnm{Kimbrough}, \binits{W.D.}},
\bauthor{\bsnm{Powell}, \binits{R.W.}}:
\batitle{Thermophysical properties of tantalum, tungsten, and tantalum-10 wt. per cent tungsten at high temperatures}.
\bjtitle{Journal of the Less Common Metals}
\bvolume{24}(\bissue{4}),
\bfpage{369}--\blpage{382}
(\byear{1971})
\doiurl{10.1016/0022-5088(71)90023-3}
\end{barticle}
\endbibitem

\bibitem[\protect\citeauthoryear{Moore et~al.}{1980}]{Moore1980_ThermalTransport_Nb}
\begin{barticle}
\bauthor{\bsnm{Moore}, \binits{J.P.}},
\bauthor{\bsnm{Graves}, \binits{R.S.}},
\bauthor{\bsnm{Williams}, \binits{R.K.}}:
\batitle{Thermal transport properties of niobium and some niobium-base alloys from 80 to 1600 k}.
\bjtitle{Journal of Applied Physics}
\bvolume{51}(\bissue{12}),
\bfpage{6308}--\blpage{6316}
(\byear{1980})
\end{barticle}
\endbibitem

\bibitem[\protect\citeauthoryear{Kresse and Hafner}{1993}]{PhysRevB.47.558}
\begin{barticle}
\bauthor{\bsnm{Kresse}, \binits{G.}},
\bauthor{\bsnm{Hafner}, \binits{J.}}:
\batitle{Ab initio molecular dynamics for liquid metals}.
\bjtitle{Phys. Rev. B}
\bvolume{47},
\bfpage{558}--\blpage{561}
(\byear{1993})
\doiurl{10.1103/PhysRevB.47.558}
\end{barticle}
\endbibitem

\bibitem[\protect\citeauthoryear{Shafeie et~al.}{2019}]{Shafeie2019}
\begin{barticle}
\bauthor{\bsnm{Shafeie}, \binits{S.}},
\bauthor{\bsnm{Guo}, \binits{S.}},
\bauthor{\bsnm{Erhart}, \binits{P.}},
\bauthor{\bsnm{Hu}, \binits{Q.}},
\bauthor{\bsnm{Palmqvist}, \binits{A.}}:
\batitle{Balancing scattering channels: A panoscopic approach toward zero temperature coefficient of resistance using high-entropy alloys}.
\bjtitle{Advanced Materials}
\bvolume{31}(\bissue{2}),
\bfpage{1805392}
(\byear{2019})
\doiurl{10.1002/adma.201805392}
\end{barticle}
\endbibitem

\bibitem[\protect\citeauthoryear{Lee et~al.}{2019}]{Lee2019}
\begin{barticle}
\bauthor{\bsnm{Lee}, \binits{S.}},
\bauthor{\bsnm{Kim}, \binits{E.M.}},
\bauthor{\bsnm{Lim}, \binits{Y.}}:
\batitle{Near‑zero temperature coefficient of resistance of hybrid resistor fabricated with carbon nanotube and metal alloy}.
\bjtitle{Scientific Reports}
\bvolume{9},
\bfpage{7763}
(\byear{2019})
\doiurl{10.1038/s41598-019-44182-7}
\end{barticle}
\endbibitem

\bibitem[\protect\citeauthoryear{Antao et~al.}{2024}]{antao2024modernizing}
\begin{botherref}
\oauthor{\bsnm{Antao}, \binits{A.}},
\oauthor{\bsnm{Burton}, \binits{J.D.}},
\oauthor{\bsnm{Dawson}, \binits{D.}},
\oauthor{\bsnm{Gemmill}, \binits{J.}},
\oauthor{\bsnm{Gerstener}, \binits{Z.}},
\oauthor{\bsnm{Godfrey}, \binits{B.}},
\oauthor{\bsnm{Groel}, \binits{S.}},
\oauthor{\bsnm{Jordan}, \binits{Z.}},
\oauthor{\bsnm{Ligon}, \binits{B.}},
\oauthor{\bsnm{Smith}, \binits{D.}}, et al.:
Modernizing clemson university's palmetto cluster: Lessons learned from 17 years of hpc administration,
1--9
(2024)
\end{botherref}
\endbibitem

\end{thebibliography}

\end{document}